# Information Acquisition and Exploitation in Multichannel Wireless Networks

Sudipto Guha* Kamesh Munagala† Saswati Sarkar‡

November 19, 2018

**Abstract**

A wireless system with multiple channels is considered, where each channel has several transmission states. A user learns about the instantaneous state of an available channel by transmitting a control packet in it. Since probing all channels consumes significant energy and time, a user needs to determine what and how much information it needs to acquire about the instantaneous states of the available channels so that it can maximize its transmission rate. This motivates the study of the trade-off between the cost of information acquisition and its value towards improving the transmission rate.

A simple model is presented for studying this information acquisition and exploitation trade-off when the channels are multi-state, with different distributions and information acquisition costs. The objective is to maximize a utility function which depends on both the cost and value of information. Solution techniques are presented for computing near-optimal policies with succinct representation in polynomial time. These policies provably achieve at least a fixed constant factor of the optimal utility on any problem instance, and in addition, have natural characterizations. The techniques are based on exploiting the structure of the optimal policy, and use of Lagrangean relaxations which simplify the space of approximately optimal solutions.

## 1 Introduction

Future wireless networks will provide each terminal access to a large number of channels. A channel can for example be a frequency in a frequency division multiple access (FDMA) network, or a code in a code division multiple access (CDMA) network, or an antenna or a polarization state (vertical or horizontal) of an antenna in a device with multiple antennas (MIMO). Several existing wireless technologies, e.g., IEEE 802.11a [1], IEEE802.11b [15], IEEE802.11h [2] propose to use multiple frequencies. For example, IEEE 802.11a protocol has 8 channels for indoor use and 4 channels for outdoor use in the 5GHz band, while the IEEE 802.11b protocol has 3 channels in the 2.4 GHz band. The potential deregulation of the wireless spectrum is likely to enable the use of a significantly larger number of frequencies. Due to significant advances in device technology, laptops with multiple antennas (antenna arrays) incorporated in the front lid, and devices with smart antennas have already been developed, and the number of such antennas are likely to significantly increase in near future.

*Department of Computer and Information Science, University of Pennsylvania, Philadelphia PA 19104. Email: sudipto@cis.upenn.edu. This research was supported in part by an Alfred P. Sloan Fellowship and NSF award CCF 04-30376.

†Department of Computer Science, Duke University. Durham NC 27708-0129. Email: kamesh@cs.duke.edu. This research was supported in part by NSF award CNS 05-40347.

‡Department of Electrical and Systems Engineering, University of Pennsylvania, Philadelphia PA 19104. Email: swati@seas.upenn.edu. This research was supported by NSF awards NCR 02-38340, CNS 04-35306, CNS 04-35141, ECCS 06-21782 and CNS 07-21308.

The increase in the number of channels is expected to significantly enhance network capacity and enable several new bandwidth-intensive applications as multiple transmissions can now proceed simultaneously in a vicinity using different channels. Furthermore, the availability of multiple channels substantially enhances the probability (at any given time) of existence of at least one channel with acceptable transmission quality, since the transmission quality of the individual channels stochastically vary with time and location of the users. These benefits can however be realized only if the users can select the channels efficiently.

Most of the existing channel selection strategies assume complete knowledge of instantaneous transmission qualities of all channels. We refer to this approach as "complete information based optimal control". Note that a user can only learn the instantaneous state of a channel by transmitting a control packet in it and subsequently the receiver informs the sender about the quality of the channel in a response packet (e.g., the RTS and CTS packet exchange in IEEE 802.11). The exchange of control packets in this probing process consumes additional energy, and prevents other neighboring users from simultaneously utilizing the channel. Probing a channel is therefore associated with a cost. When the number of available channels is large, the cost incurred in learning the instantaneous transmission qualities of all channels may become prohibitive. Owing to this cost, some recent papers have investigated selection strategies that assume no knowledge of instantaneous transmission qualities of any channel [27]. This approach, which we refer to as "minimal information based optimal control", may however attain significantly lower transmission rates owing to sub-optimal selection of channels. We seek to design a framework that attains any desired trade-off between the above extremes using only simple control mechanisms. Specifically, we develop a framework for *partial information based stochastic control* which, in accordance with the costs and the benefits of probing different channels, determines both (a) the amount of information a user must obtain about the instantaneous transmission qualities of the channels at its disposal and also (b) how to select the channels based on the acquired information.

We consider a single sender with access to $n$ channels. Every time the sender probes a channel it learns about the signal to noise ratio and thereby the probability of success in the channel, but also incurs a certain cost which may again be different for different channels. Before each transmission, the sender needs to determine how many and which channels it will probe and also the sequence in which these channels will be probed (*probing policy*). In this paper, we consider the scenario where a sender can transmit in only one channel in a time slot and transmits at most one packet in each slot. Based on the outcomes of the probes, a sender decides whether to transmit or defer transmission until transmission qualities improve (*transmission policy*). If the sender decides to transmit, it must select one of the available channels (*channel selection policy*), which need not be those that it has probed. We seek to design a jointly optimal probing and channel selection policy that maximizes a system utility which is the difference between the probability of successful transmission and a suitably scaled expected probing cost before each transmission. Loosely, this utility function represents the gain or the profit of the sender if the sender receives credit from the receiver for each packet it delivers successfully and needs to additionally compensate the wireless provider for each probe packet it transmits.

**Technical Hurdles and Contributions:** The optimal probing policy needs to probe adaptively, i.e., the result of a probe determines the channels to be probed subsequently. For example, consider channels with 3 possible states $(0, 1, 2)$, each of which is associated with a different transmission quality. Clearly, the probing terminates if a probed channel is in the highest state. Now, let a probed channel be in the intermediate state (state 1). Then the subsequent probes should be limited to channels that have high probabilities of being in the highest state. However, if all channels that have been probed in a slot are in the lowest state, then the channels that have high probabilities of being in the intermediate state may also be subsequently probed. Even the decision regarding which channel should be probed first depends on the high order statistics of *all* the channels in a complex fashion. This is because the optimum policy may not probe a channel if its quality has a low variance as probing it does not provide significant information but incurs additional cost. Example 5.1 in Section 5 illustrates these points. Also, the channel selection decision depends on the outcomes of the probes and the expectation and uncertainty of the transmission quality of the channels that have not been probed. The optimal policy is therefore a decision tree over $n$ variables (Figure 1) and can be computed by solving a dynamic program with $\Omega(K2^n)$ states – naive computations will require both exponential time and exponential storage space. The above observations rule out greedy strategies for computing the optimal solution. Our main contribution is in showing that despite these hurdles, there is a nice (albeit non-trivial) combinatorial structure in the optimal decision trees. We then use this structure to design simple, natural, and polynomial time greedy algorithms which provably achieve at least $4/5$ of the optimal utility on any problem instance.

A nice feature of our algorithmic framework is that it easily extends to handle other constraints on the problem, as we elucidate next. For example, when the sender is not saturated, that is it does not always have packets to transmit, it need not transmit packets in every slot, and therefore needs to jointly optimize the transmission, probing and channel selection policy so as to attain the maximum utility. This presents additional technical challenges since the transmission policy needs to take into account two conflicting criteria: the transmissions should only happen when some channel is observed to be in a very high quality state, but on the other hand, they should happen frequently enough to maintain stability. This leads to a rate constraint in the corresponding optimization. We show a novel technique based on linear programming duality to handle the rate constraint, and present simple polynomial time computable greedy policies which provably approximate the optimal policies. In addition, these greedy policies are the most natural threshold policies, where the decision is to transmit only when the reward from transmission exceeds a certain threshold. This threshold depends on the arrival rate, and is computed by a simple parametric search.

**Summary of Contributions:** In summary, our main contribution is to obtain succinct polynomial time computable joint probing, selection and transmission policies that provably attain utility values which are within constant factors of the optimal utilities. More specifically:

1. We first consider the case that a sender is *saturated* and seek to determine the policy that has the maximum utility. We prove that when each channel has two states an optimal policy can be computed in $O(n \log n)$ time (Section 4). When each channel has $K$ states, we obtain a policy that provably attains $4/5$ of the maximum utility and can be computed in $O(n^2K)$ time (Section 5). These performance guarantees hold even when different channels have different distributions for the state processes and different probing costs. In the special case in which all channels have equal probing costs, but potentially different distributions for the state processes, we present a parametrized probing and channel selection policy whose parameters can be appropriately selected so as to attain any desired trade-off between performance guarantee and computation time (Section 6).

2. We next consider the *unsaturated* sender scenario where packets arrive with a given rate $\lambda$, so that the sender will not have packets to transmit in every slot (Section 7). The goal now is to determine the policy that attains the maximum utility among all stable policies. We prove that when each channel has two states such a policy can be computed in $O(n^3)$ time and in the case where each channel has $K$ states, we show that a stable policy that provably attains $2/3$ of the maximum utility among all stable policies can be computed in $O(n^2K(n+K))$ time.

All policies can be readily implemented in resource constrained devices as once computed they can be executed in $O(n)$ time and stored in $O(n+K)$ space.

Our results are somewhat surprising given that optimal solutions for most partial information based control problems are possibly computationally intractable, and standard approximation techniques either do not provide guaranteed approximation ratios or require exponential computation times [5]. Our proofs therefore rely on exploitation of specific system characteristics and employ techniques that are not standard in context of stochastic control. The techniques we develop are very natural and general; they are expected to have wider applicability in designing simple and intuitive heuristics for a larger class of problems in the broad area of partial information based control problems, and in particular the joint optimization of the reward obtained from informed selections and the cost incurred in acquiring the required information. We will explore this in future work.

## 2 Related Literature

We first discuss the relation of our problem with some classical problems like the stopping time and multi-armed bandit problems. The most well-researched version of the stopping time problem is a stochastic control problem that optimally selects between two possible actions at any given time: to continue or to stop [10]. Recently, the results for this problem have been used to solve partial information based control problems for statistically identical channels with equal probing costs [19, 24]. Empirical investigations indicate that different channels available to a sender oftentimes have different statistics [16]. When channels have different statistics and/or different probing costs, which is the case we consider in the paper, the optimal action needs to be selected from multiple options at any given time - the options being (a) whether to continue probing (b) which channel to probe next if the decision is to probe and (c) which channel to transmit if the decision is to stop probing. Thus, the results from the above version of stopping time problem do not apply in our context. The optimal stopping time problem has also been considered in a more general setting where the number of available actions may be more than two; our problem is in fact a special case of this general version (Chapter IV, [5]). In this general case, the process terminates in certain states, which constitute the termination set, and selects the optimal action in other states. But, so far, only certain broad characterizations of the termination set are known in this general case, and the optimal actions when the decision is not to stop are also not known in closed form [5]. Thus, these general results do not lead to the optimal policies we are seeking to characterize.

The stochastic multi-armed bandit problem considers a bandit with $n$ arms [14]. The system can try one arm in each slot, and when it tries an arm, it receives a random reward which depends on the state of the arm. The state of an arm changes only when the system tries it. The reward of a system in $T$ slots is the sum of the rewards in each slot. The goal is to maximize the expected reward in $T$ slots. Our problem differs from the above in that (a) the state of a channel can change even when it is not probed or used for transmission and (b) a node can learn the states of multiple channels in an epoch while incurring additional probing costs for learning the state of each additional channel and it can choose each such channel adaptively after observing the states of the channels chosen for probing before in the same slot. The adversarial multi-armed bandit problem [3] and the restless bandit problem [6, 28] remove one of the above differences in that they allow the state of an arm to change even when the system does not try it. But, the adversarial multi-armed bandit problem [3] seeks to optimize the selection under the assumption that the sender uses the same arm in all slots. Note that we allow a sender to probe, and also transmit, in different channels in different slots. In another version of the adversarial multi-armed bandit problem, the goal is to select the arms so as to minimize the "regret" or the difference in expected reward with the best policy in a collection of a certain number of given policies [3]. Our problem differs from this version of adversarial multi-armed bandit problem and also from the restless bandit problem [6, 28] in that we allow a node to adaptively probe different channels in the same slot by paying additional costs (difference (b) above). Thus the results available in this context do not apply in our problem, and we use different solution approach and obtain different performance guarantees.

Optimizing the order of evaluation of random variables so as to minimize the cost of evaluation ("pipelined filters") has been investigated in several different contexts like diagnostic tests in fault detection and medical diagnosis, optimizing conjunctive query and joint ordering in data-stream systems, web services [4, 25, 11, 13, 17, 20, 21, 22, 23]. However our work is different from all the above in that, we (a) consider multi-state channel models whereas pipeline filters consider two state models and (b) allow a node to transmit in a channel even if the channel has not been probed. Note that usually two state models can not capture the statistical variations of wireless channels [16]. As we demonstrate later, both the above generalizations significantly alter the decision issues and the optimal solutions (Section 5).

Finally, opportunistic selection of channels with complete knowledge of channel states has been comprehensively investigated over the last decade (e.g., [26]). But, in general, the area of partial information based control problems, and in particular the joint optimization of the reward obtained from informed selections and the cost incurred in acquiring the required information, remains largely unexplored in wireless networks. Policies with provable performance guarantees are known only in special cases like statistically identical channels [8, 19, 24], and even under these simplifying assumptions only the saturated sender case had been investigated. We consider both the saturated and unsaturated sender case, and in both cases obtain provable performance guarantees even when channels have different statistics and/or probing costs. The results in this paper therefore enhance the state of knowledge in an emerging area which has hitherto received only limited attention.

# 3 System Model and Problem Definition

A sender $U$ has access to $n$ channels which are denoted as channels $1, 2, \ldots, n$, each of which has $K$ possible states, $0, \ldots, K-1$. We assume that time is slotted. In any slot channel $j$ is in state $i$ with probability $p_{ij}$ independent of its state in other time slots and the states of other channels in any slot. Without loss of generality, we assume that $p_{K-1j} < 1$ for each $j$, as otherwise the optimum policy is simply to transmit in $j$ without probing any channel. In every slot, $U$ transmits at most one data packet in a selected channel. If the channel selected for transmission is in state $i$, the transmission is successful with probability $r_i$. Without loss of generality we assume $0 \leq r_0 < r_1 < \cdots < r_{K-1}$. For simplicity, we also assume that $r_0 = 0$; all analytical results can however be generalized to the scenario where $r_0 > 0$. Whenever $U$ probes a channel $j$, it pays a cost of $c_j \geq 0$. Probing different channels may incur different costs as the probing process for different channels may interfere with the channel access of different number of users (based on geometry and allocation of channels). We now formally define the policies and the performance metrics.

**Definition 3.1.** *A* **probing policy** *is a rule that, given the set of channels the sender has already probed in a slot (which would be empty at the beginning of the slot) and the states of the channels probed in the slot, determines (a) whether the sender should probe additional channels and (b) if the sender probes additional channels which channel it should probe next. The sender knows the state of a channel in a slot if and only if it probes the channel in the slot.*

**Definition 3.2.** *A* **selection policy** *is a rule that selects a channel for the transmission of a data packet in a slot on the basis of the states of the probed channels, after the completion of the probing process in the slot. The selection policy can select a channel even if it has not been probed in the slot, and in that case, the channel is referred to as a* **backup** *channel.*

**Definition 3.3.** *The* **probing cost** *is the sum of the costs of all channels probed in the slot. The probing cost is clearly a random variable that depends on the probing policy and the outcomes of the probes (as the sender may probe subsequent channels depending on the outcomes of the previous probes). The* **expected probing cost** *is the expectation of this random variable and depends on both the probing policy and the channel statistics.*

**Definition 3.4.** *In any slot, the* **transmission reward** *is* $1$ *if there is a successful transmission and* $0$ *otherwise. Therefore, the expected transmission reward is* $r_i$ *in a slot* $t$ *if* $U$ *transmits in a channel in state* $i$ *during $t$. The expected transmission reward of a policy is therefore $\sum_i q_i r_i$ where $q_i$ is the probability that the selection policy decides to use a channel which is in state $i$; $q_i$ depends on the channel statistics as well as the policy.*

**Definition 3.5.** *The* **expected utility** *of the sender, denoted simply as* **gain***, is the difference between the expected transmission reward, and the probing cost scaled by a factor $\kappa$. We denote the gain of a policy $\pi$ as $G_\pi$.*

The gain depends on the probing and selection policies, the channel statistics and the scaling parameter $\kappa$ – choosing the scaling parameter $\kappa$ to be 0 makes the policy acquire complete information, while setting it to $\infty$ makes the policy acquire no information. Since $\kappa$ can be included in the probing costs themselves, we drop this parameter in the remaining discussion without loss of generality.

In Sections 4, 5 and 6, we assume that $U$'s queue is never empty (saturated sender assumption); we relax this assumption in Section 7. The two versions of the problem are defined as follows:

**Saturated Sender Problem:** Under the saturated sender assumption, at least one policy that maximizes the utility transmits a packet in every slot. We therefore assume that $U$ transmits exactly one data packet in every slot. The problem formulation for the saturated sender case follows.

**Problem 1.** *Given $\{c_j\}$, $\{r_i\}$ and $\{p_{ij}\}$, find a probing and selection policy so as to maximize the expected gain. Let* OPT *denote the optimal policy and $G_{Opt}$ its gain.*

Since channels are temporally independent, the optimal policy in a slot need not depend on the decisions and the observations in other slots. Also, the optimal policy remains the same in all slots, though the specific choices it makes may be different in different slots depending on the outcome of the probes. Note that the optimal probing policy does not probe any further in a slot if a probed channel is in state $K-1$. Using these observations, the optimal policy can be computed using a bottom-up dynamic program whose states correspond to the tuple consisting of (a) the maximum value of the best state encountered so far and (b) the set of channels that have not been probed yet. Thus, the dynamic program has $K2^n$ states, and hence, naive computations will require $\Omega(K2^n)$ time and space.

**Policies and Decision Trees:** Every joint policy can be represented by a unique decision tree (Figure 1); we therefore use policies and decision trees interchangeably.

**Unsaturated Sender Problem:** We now relax the assumption that a sender always has packets to transmit. A sender generates packets as per an arrival process which constitutes a positive recurrent, aperiodic, irreducible Markov chain. Under the steady state distribution of the Markov chain, the expected number of arrivals in any slot is $\lambda$, where $\lambda \in [0, 1)$. Packets are stored in an infinite buffer. If in a slot the sender has a packet in its queue, the slot is referred to as a *busy* slot. The sender transmits only in busy slots, but may not transmit in every busy slot; it may improve its gain by deferring transmission until at least one channel has good quality. The *transmission policy* is a rule that determines which slots a sender transmits. The decisions may depend on the outcomes of the probes, the queue lengths, channel and arrival statistics. The sender must however ensure that it transmits at least at the rate at which it generates packets, otherwise its delay becomes unbounded. In addition to gain, system stability is therefore of interest.

**Definition 3.6.** *The system is stable if the sender's expected queue length is finite. A policy that attains finite expected queue length is a stable policy.*

**Problem 2.** *Given $\{c_i\}$, $\{r_i\}$ and $\{p_{ij}\}$ find a probing, selection and transmission policy that stabilizes the system and maximizes the gain among all stable policies. Let* OPTUNSAT *denote the optimal policy and $G_U$ its gain.*

## 4 The Two-State, Saturated Sender Problem

We now consider the case that the sender always has packets to transmit and seek to solve **Problem** 1 formulated in Section 3 when $K = 2$. We first consider a specific class of policies, EXHAUST, and prove that OPT belongs in this class. Subsequently we show how to find the optimal policy TWOSTATEOPT in this class in $O(n \log n)$ time. Thus, when $K = 2$, TWOSTATEOPT is OPT and can be computed in $O(n \log n)$ time. Also, TWOSTATEOPT can be executed and stored in $O(n)$ time and space.

**Definition 4.1.** *Given $S \subset \{1, \ldots, n\}, i \notin S$, let* EXHAUST*$(S, i)$ denote the class of policies which probe all channels in $S$ in a deterministic order until a probed channel is in state $1$ or all channels in $S$ have been probed. It selects the last probed channel if it is in state $1$, and selects $i$ otherwise. Channel $i$ is denoted as the backup channel.*

In what follows, we prove that there is an optimal policy which is of the form EXHAUST$(S, i)$. We note that results proved later in the paper (for the case $K > 2$) will subsume the proof of this fact – but a straightforward application of these later results will yield an algorithm which requires $O(n^2)$ time.

**Lemma 4.1.** *There exists an* EXHAUST$(S, i)$ *policy which is optimal.*

*Proof.* We prove the lemma by induction on the number of channels $n$. For the base case, $n = 1$, the expected gain is $r_1 p_{11} - c_1$ if the optimal policy probes the channel, and $r_1 p_{11}$ otherwise. Since $c_1 \geq 0$, the policy that selects the channel without probing is optimal over all possible convex combinations, i.e., randomization, of the above two policies. Thus, EXHAUST$(\Phi, 1)$ is an optimal policy in this case.

Assuming the lemma holds for $n = s$, consider a set $J$ of $s + 1$ channels. OPT can either (a) select a channel without probing or (b) probe a channel. Conditioned on case (a), $G_{\text{Opt}} = G_{\text{EXHAUST}(\Phi,j)}$, where $j = \arg\max_i p_{1i}$. In case (b), OPT chooses to probe a channel $i$ with some probability. Subsequently, if $i$ is in state 1, OPT selects $i$. Now, if $i$ is in state 0, then OPT takes the same decisions as that in a system with the $s$ remaining channels, and by the induction hypothesis, uses EXHAUST$(Q, j)$ policy for some $Q \subset J \setminus \{i\}, j \in (J \setminus Q) \setminus \{i\}$. Thus, in this case, OPT is an EXHAUST$(\{i\} \circ Q, j)$ policy where the $\circ$ denotes the ordering. Therefore, conditioned on case (b), $G_{\text{Opt}}$ is a convex combination of the gains of EXHAUST policies. Therefore, overall, $G_{\text{Opt}}$ is a convex combination of the gains of EXHAUST policies. Thus, there exists an optimum policy which is EXHAUST$(S, i)$. □

We next prove that OPT satisfies additional properties, which allows a polynomial time computation of OPT.

**Lemma 4.2.** *Let $S_i = \{j : (1 - p_{1i}) p_{1j} r_1 > c_j\}$. If* EXHAUST$(S, i)$ *is an optimum policy, then the following conditions hold.*

1. *channels $j$ in $S$ are probed in decreasing order of $p_{1j}/c_j$*
2. EXHAUST*$(S_i, i)$ policy is an optimum policy.*

*Proof.* Let EXHAUST$(S, i)$ policy be an optimum policy. Wlog. $S = \{k_1, \ldots k_{|S|}\}$, where channel $k_l$ is probed before $k_{l+1}$. Then the gain of EXHAUST$(S, i)$ policy is

$$A = \sum_{l=1}^{|S|} (p_{1k_l} r_1 - c_{k_l}) \prod_{m=1}^{l-1} (1 - p_{1k_m}) + p_{1i} r_1 \prod_{m=1}^{|S|} (1 - p_{1k_m}).$$

We first prove (1). Recall that $p_{1j} < 1$ for all channels $j$. Let $p_{1k_s}/c_{k_s} < p_{1k_{s+1}}/c_{k_{s+1}}$. Consider a new policy which probes $k_{s+1}$ before $k_s$ but is other-wise similar to the EXHAUST$(S, i)$ policy. Let the gain of this new policy be $B$. Then, $A - B = \prod_{m=1}^{s-1}(1 - p_{1k_m})(p_{1s}c_{s+1} - p_{1s+1}c_s)$. Thus, clearly, $B > A$. Thus, EXHAUST$(S, i)$ is not the optimum policy. The result follows.

We now prove (2). If $S = S_i$ the result follows. Let $S \neq S_i$. Thus, either $S_i \setminus S \neq \phi$ or $S \setminus S_i \neq \phi$.

Let $S \backslash S_i \neq \phi$. Consider some $j \in S \backslash S_i$. From (1), $p_{1k_s}/c_{k_s} \geq p_{1k_{s+1}}/c_{k_{s+1}}$. Thus, $(1-p_{1i})p_{1k_{|S|}}r_1/c_{k_{|S|}} = \min_{1 \leq l \leq |S|}(1 - p_{1i})p_{1l}r_1/c_l \leq (1 - p_{1i})p_{1j}r_1/c_j \leq 1$. Thus, $k_{|S|} \in S \setminus S_i$. Let $Q = S \setminus \{k_{|S|}\}$. The gain of EXHAUST$(Q, i)$ policy with probing sequence $k_1, \ldots, k_{|S|-1}$ is $D = \sum_{l=1}^{|S|-1}(p_{1k_l}r_1 - c_{k_l})\Pi_{m=1}^{l-1}(1 - p_{1k_m}) + p_{1i}r_1\Pi_{m=1}^{|S|-1}(1 - p_{1k_m})$. Now, $D - A = (c_{k_{|S|}} - (1 - p_{1i})p_{1k_{|S|}}r_1)\Pi_{m=1}^{|S|-1}(1 - p_{1k_m})$. Since $(1 - p_{1i})p_{1k_{|S|}}r_1 \leq c_{k_{|S|}}$, $D \geq A$. Thus, EXHAUST$(Q, i)$ is an optimum policy, where $Q \subseteq S$ and $|Q \setminus S_i| < |S \setminus S_i|$. Continuing this argument, clearly there exists a $T$ such that $T \subseteq S$ and $T \setminus S_i = \phi$ and EXHAUST$(T, i)$ policy is optimal.

Now let $S_i \setminus S \neq \phi$. If $S \setminus S_i \neq \phi$, let $T$ be as constructed in the above paragraph; otherwise let $T = S$. In both cases, EXHAUST$(T, i)$ policy is optimal. We now show that $S_i \setminus T = \phi$. If not, consider a $j \in S_i \setminus T$. Let $Q = T \cup \{j\}$. The gain of EXHAUST$(Q, i)$ policy with probing sequence $k_1, \ldots k_{|T|}, k_j$ is $C = \sum_{l=1}^{|T|}(p_{1k_l}r_1 - c_{k_l})\Pi_{m=1}^{l-1}(1 - p_{1k_m}) + (p_{1j}r_1 - c_j)\Pi_{l=1}^{|T|}(1 - p_{1k_l}) + p_{1i}r_1(1 - p_{1j})\Pi_{l=1}^{|T|}(1 - p_{1k_l})$. Now, $C - A = ((1 - p_{1i})p_{1j}r_1 - c_j)\,\Pi_{l=1}^{|T|}(1 - p_{1k_l})$. Since $p_{1s} < 1$ for all $s$ and $(1 - p_{1i})p_{1j}r_1 > c_j$, $C > A$. This contradicts the optimality of the EXHAUST$(T, i)$. Thus, $S_i \setminus T = \phi$. Thus, $S_i = T$. Hence, EXHAUST$(S_i, i)$ policy is optimal.

□

Lemmas 4.1 and 4.2 prove that there exists an EXHAUST$(S_i, i)$ policy that is optimal, and this policy probes the channels in $S_i$ in decreasing order of $p_{1j}/c_j$. The routine DETERMINE BEST BACKUP described below determines the BEST BKUP channel $i^*$ such that EXHAUST$(S_{i^*}, i^*)$ attains the maximum gain among all such EXHAUST$(S_i, i)$ policies. Note that $i^*$ can be computed in $O(n^2)$ time using a naive implementation, but the following computation requires only $O(n \log n)$ time.

DETERMINE BEST BACKUP

1. Sort the channels so as to arrange them in decreasing order of $p_{1i}/c_i$. Re-number the channels in accordance with the sorted order, i.e., if $i < j$, $p_{1i}/c_i > p_{1j}/c_j$. Let $S_i = \{j : (1 - p_{1i})p_{1j}r_1 > c_j, j \neq i\}$. Let Gain$(i)$ denote the gain of EXHAUST$(S_i, i)$.
2. Let $D_0 = 1$. For $j \geq 0$, compute $D_{j+1} = D_j(1 - p_{1j})$.
   /* $D_{j+1}$ = Probability that first $j$ channels return state "0" */
3. Let $F_0 = 0$. For $j \geq 1$, compute $F_{j+1} = F_j + (p_{1j}r_1 - c_j)D_j$.
   /* $F_{j+1}$ = Gain of the first $j$ channels if probed. */
4. For each channel $i$, if $i > |S_i|$, Gain$(i) = F_{|S_i|+1} + p_{1i}r_1D_{|S_i|+1}$;
   else Gain$(i) = F_i + \frac{F_{|S_i|+2} - F_i - (p_{1i}r_1 - c_i)D_i}{1 - p_{1i}} + \frac{p_{1i}r_1}{1 - p_{1i}}D_{|S_i|+2}$.
5. Let BEST BKUP $= \arg\max_{i=1,\ldots,n}$ Gain$(i)$.

We now explain the computations of the gains of EXHAUSTpolicies in the DETERMINE BEST BACKUP routine. The channels are numbered in decreasing order of $p_{1j}/c_j$. Now, $F_{|S_i|+1}$ is the gain of sequentially probing the first $|S_i|$ channels, and $D_{|S_i|+1}$ is the probability that the first $|S_i|$ channels are in state 0. If $i > |S_i|$, then the first $|S_i|$ channels constitute $S_i$. Thus, the gain of EXHAUST$(S_i, i)$ is $F_{|S_i|+1} + p_{1i}D_{|S_i|+1}$. If $i \leq |S_i|$, the first $|S_i| + 1$ channels constitute $S_i \cup \{i\}$. Thus, $F_i + \frac{F_{|S_i|+2} - F_i - (p_{1i}r_1 - c_i)D_i}{1 - p_{1i}}$ is the gain obtained by probing channels in $S_i$ in decreasing order of $p_{1j}/c_j$, and $D_{|S_i|+2}/(1 - p_{1i})$ is the probability that all channels in $S_i$ are in state 0. Thus, the gain of EXHAUST$(S_i, i)$ is $F_i + \frac{F_{|S_i|+2} - F_i - (p_{1i}r_1 - c_i)D_i}{1-p_{1i}} + \frac{p_{1i}r_1}{1-p_{1i}} D_{|S_i|+2}$. The computation time for the DETERMINE BEST BACKUP routine is dominated by the time required to sort the channels, which is $O(n \log n)$.

The following TWOSTATEOPT policy is EXHAUST$(S_{i^*}, i^*)$ where $i^*$ is the BEST-BKUP channel returned by the routine DETERMINE BEST BACKUP.

> TWOSTATEOPT
>
> 1. Probe channels $j \in S_{\text{BEST-BKUP}}$ until a probed channel is in state 1 or all channels in $S_{\text{Best-Bkup}}$ have been probed.
> 2. If the last probed channel $j$ is in state 1, transmit the packet in $j$, else transmit the packet in the BEST-BKUP channel.

**Theorem 4.3.** TWOSTATEOPT *attains the maximum gain when $K = 2$, and can be computed in $O(n \log n)$ time.*

Since TWOSTATEOPT is the EXHAUST$(S_{i^*}, i^*)$ policy that attains the maximum gain among all EXHAUST$(S_i, i)$ policies that probe the channels in $S_i$ in decreasing order of $p_{1j}/c_j$, its optimality follows from Lemmas 4.1 and 4.2. The computation time for TWOSTATEOPT is the same as that for the routine DETERMINE BEST BACKUP which can be computed in $O(n \log n)$ time. Thus, Theorem 4.3 follows. Clearly, TWOSTATEOPT can be executed and stored in $O(n)$ time and space.

# 5 The Multi-State Saturated Sender Problem

We still assume that the sender always has packets to transmit but now focus on the case that each channel has $K$ states where $K > 2$. We first demonstrate that some natural generalizations of the TWOSTATEOPT policy are suboptimal when $K > 2$.

**Example 5.1.** *Recall that* TWOSTATEOPT *probes channels in increasing order of $c_j/p_{1j}$. Thus, for $K > 2$, the natural generalizations of* TWOSTATEOPT *are to probe channels in decreasing order of the ratio between their (a) probabilities of being in the highest state and costs (i.e., $p_{K-1j}/c_j$) or (b) the expected rewards and costs (i.e., $\sum_{k=0}^{K-1} p_{kj} r_j / c_j$). Figure 1 presents one scenario where both these probing sequences are sub-optimal. Note that $i$ has the least, $j$ intermediate and $k$ maximum expected rewards, and $p_{2k}/c_k > p_{2j}/c_j > p_{2i}/c_i$. But,* OPT *probes $i$ before probing $j$.*

The main challenge for $K > 2$ is that the optimal probing sequence needs to be adaptively determined depending on the outcomes of the previous probes in a slot (Figure 1). For example, when $K = 3$, and when a probed channel is in the intermediate state (state 1), then the subsequent probes should be limited to channels that have high probabilities of being in state 2. However, if all channels that have been probed in a slot are in state 0, then the channels that have high probabilities of being in state 1 may also be subsequently probed.

We show that in $O(n^2 K)$ time, we can compute a policy which attains $4/5$ of the maximum gain, for arbitrary distributions for the state processes and costs. However, more importantly, we develop techniques and ideas, such as the *Structure Theorem* below, which are useful beyond the context of this specific problem. In fact, we will use the structure theorem for all the problems considered in the rest of this paper.

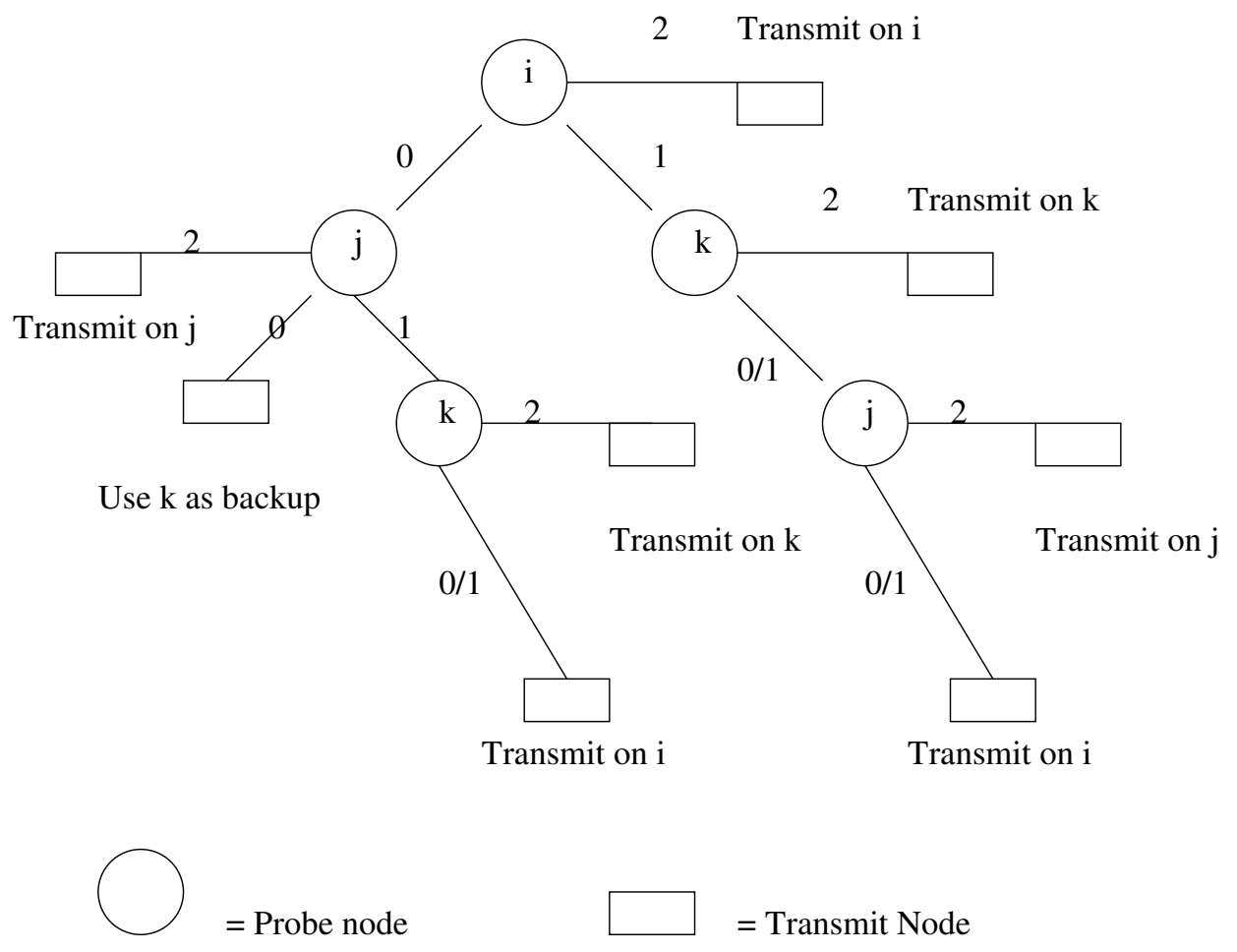

Figure 1: Consider a node $U$ that has access to 3 channels $i, j, k$ each of which has 3 states. Let $r_2 = 1, r_1 = 0.1$. The probabilities associated with different states of $i, j, k$ are $(0.49, 0.02, 0.49)$, $(0.5, 0.01, 0.49)$, $(0.5, 0.5 - \delta, \delta)$ respectively. Also, $c_i = 0.05885\delta, c_j = 0.06\delta, c_k = 0.05\delta$. Let $\delta < 0.15$. The figure shows the decision tree for OPT in this case. A channel is probed at each probe node, and the letter inside it indicates which channel is probed at the node. The numbers next to the branches indicate the outcome of the probe. The number $r/s$ next to a branch indicates that both states $r$ and $s$ of the previously probed channel lead to the same action. For example, the sender first probes channel $i$. If $i$ is in state 2, it transmits in $i$. If $i$ is in state 1 and 0, it probes $k$ and $j$ respectively.

## 5.1 The Roadmap and the Main Results

The main component of the algorithm is the following theorem which is proved in Section 5.2.

**Theorem 5.1** (Structure Theorem)**.** *There exists an optimum policy which uses a unique backup channel.*

The theorem states that there exists an optimum policy OPT and a channel $\ell$ such that whenever OPT uses a backup, it uses $\ell$ as a backup. Note that the above is true for the case $K = 2$ trivially, because if at any point the sender observes a channel to be in state 1, there is no further benefit of probing. Thus the strategy corresponds to a path where we observe every probed channel to be in state 0. Note that another interesting property of the case $K = 2$ is that the backup channel is never probed. This motivates the following definitions.

**Definition 5.1.** *Let $\mathcal{P}(\ell)$ denote the class of policies, each of which (a) never probes $\ell$ and (b) never uses any channel other than $\ell$ as a backup. Let $\mathcal{P}(0)$ correspond to the class of policies that never use backup channels (note that the channels are numbered $1, 2, \ldots, n$). Let the policy that attains the maximum gain among all policies in $\mathcal{P}(\ell)$ be denoted as* RESERVEBKUP($\ell$)*, and the policy that attains the maximum gain among all policies in $\cup_{l=0}^{n} \mathcal{P}(\ell)$ be denoted as* BESTRESERVEBKUP*. Let $G_{BestReserveBkup}$ be the gain of* BESTRESERVEBKUP*.*

The following theorems indicate why BESTRESERVEBKUP is of interest.

**Theorem 5.2.** *For $\ell = 0, 1, \ldots, n$, we can compute a policy* RESERVEBKUP($\ell$) *that attains the maximum gain among all policies in $P(\ell)$ in time $O(nK \log n)$.*

RESERVEBKUP($\ell$) has been presented in Section 5.3, and the above theorem has been proved in Section 5.4. Therefore, clearly, we can compute BESTRESERVEBKUP in time $O(n^2 K \log n)$. In Section 5.3, we argue that we can in fact compute BESTRESERVEBKUP in time $O(n^2 K)$.

**Theorem 5.3.** $G_{BestReserveBkup} \geq (4/5)G_{Opt}$.

*Proof.* By the structure Theorem (Theorem 5.1), there exists an optimum policy OPT that uses a unique backup, say $B$. Let $\alpha$ denote the probability with which OPT uses $\ell$ as a backup. Construct a new policy $A$ that is similar to OPT except that it probes $B$ whenever OPT uses $B$ as a backup. Clearly, $A$ attains a gain of at least $G_{\text{Opt}} - \alpha c_B$. Also, since $A \in \mathcal{P}(0)$, its gain is at most $G_{\text{BestReserveBkup}}$. Thus,

$$G_{\text{BestReserveBkup}} \geq G_{\text{Opt}} - \alpha c_B. \tag{1}$$

We now show that there exists a policy which never probes $B$, but does not perform significantly worse than OPT. In this discussion, the gain $G(T)$ of a sub-tree $T$ rooted at $t$ is defined as the expected reward owing to transmissions at the leaves of $T$ conditioned on reaching $t$, minus the expected probing cost of nodes in $T$.

Suppose that OPT probes $B$ at nodes $m_1, \ldots, m_J$ in its decision tree. Let $\beta_1, \ldots, \beta_J$ be the respective probabilities that OPT traverse these nodes and $G_1, \ldots, G_J$ be the respective gains of OPT given that it traverses these nodes. Now consider the gains of the subtrees $G'_1, \ldots, G'_J$ produced by modifying the decision tree so that $B$ is not probed. This produces a decision tree $\tau$ which is the same as that for OPT except for the trees rooted at $m_1, \ldots, m_J$. $\tau$ reaches these nodes with the same probabilities as OPT. We now make a claim which allows us to complete the proof, and subsequently we prove why the claim is true. This claim would also be used for other results.

**Claim 5.4.** *Let $T$ be any decision (sub-)tree rooted at node $t$ where we probe channel $u$, and let its gain be $G(T)$. Suppose at the point of arriving at decision node $t$, the best probed channel has state at least $j \geq 0$ (equivalently, reward at least $r_j$). Then there exists a corresponding (sub-)tree $T'$, where $u$ is not probed at $t$ or anywhere else in $T'$, and whose gain $G(T')$ satisfies $G(T') \geq G(T) + c_u - \sum_{i=j+1}^{K-1} p_{iu} r_i$.*

We first complete the proof of the theorem using the above Claim 5.4. Setting $j = 0$ it follows that $G_k - G'_k \leq \sum_{i=0}^{K-1} p_{iB} r_i - c_\ell$. Then the difference between the overall gains of OPT and $\tau$ is $\sum_{k=1}^{J} \beta_k (G_k - G'_k)$. Since $\sum_{t=1}^{J} \beta_t \leq 1 - \alpha$, the policy $\tau$ which never probes $B$, attains a gain of at least $G_{\text{Opt}} - (1 - \alpha)(\sum_{i=0}^{K-1} p_{iB} r_i - c_B)$.

Since $\tau$ never probes $B$, therefore $\tau \in \mathcal{P}(B)$. Thus we have,

$$G_{\text{BestReserveBkup}} \geq G_{\text{Opt}} - (1 - \alpha)(\sum_{i=0}^{K-1} p_{iB} r_i - c_B). \tag{2}$$

From multiplying Equations (1) and (2) with $(1-\alpha)$ and $\alpha$ respectively, and adding the results, we have $G_{\text{BestReserveBkup}} \geq G_{\text{Opt}} - \alpha(1-\alpha) \sum_{i=0}^{K-1} p_{iB} r_i$. Now, the policy that uses $B$ as a backup without probing any channel attains a gain of $\sum_{i=0}^{K-1} p_{iB} r_i$. Since this policy is in $\mathcal{P}(B)$, $\sum_{i=0}^{K-1} p_{iB} r_i \leq G_{\text{BestReserveBkup}}$. Thus, $G_{\text{BestReserveBkup}} \geq G_{\text{Opt}}/(1 + \alpha(1-\alpha))$. The result follows since the maximum value of the denominator is 1.25. This proves the theorem; we now focus on Claim 5.4.

*(Proof of Claim 5.4).* Let $F_{iu}$ denote the set of leaf nodes in $T$ where the decision is to transmit on probed channel $u$ in state $i$ (note that this happens only if $i > j$). Let $W$ be the event that $t$ is reached, and $\hat{F}_{iu}$ be the event that $F_{iu}$ is reached. First note that $\Pr[\hat{F}_{iu}|W] \leq p_{iu}$ for all states $i$. Now construct a corresponding decision tree $T''$ in which the decision at $F_{iu}$ is to transmit on the (a) best probed channel excluding the probed channel $u$ if channels other than $u$ have been probed and (b) a backup channel otherwise. Clearly,

$$G(T'') \geq G(T) - \sum_{i=j+1}^{K-1} \Pr[\hat{F}_{iu}|W] r_i \geq G(T) - \sum_{i=j+1}^{K-1} p_{iu} r_i$$

In the decision tree $T''$, the transmission is never on probed channel $u$. Therefore, at the root node $t$ in $T''$ where $u$ is probed, suppose we do not actually probe $u$ (saving cost $c_u$), but simply choose the branch corresponding to $u$ being in state $i$ with probability $p_{iu}$, this new decision tree $T^*$ has gain:

$$G(T^*) \geq G(T'') + c_u \geq G(T) - \sum_{i=j+1}^{K-1} p_{iu} r_i + c_u$$

This new decision tree neither probes nor uses $u$. Denote the sub-tree of $T''$ below $t$ corresponding to channel $u$ being in state $i$ to be $T_i$. If $i^* = \text{argmax}_{i=1}^{K} G(T_i)$, we have:

$$G(T^*) = \sum_{i=1}^{K} p_{iu} G(T_i) \leq \max_{i=1}^{K} G(T_i) = G(T_{i^*})$$

Modify $T^*$ so that on reaching node $t$, branch $T_{i^*}$ is chosen. Denote this $T'$. Then,

$$G(T') = G(T_{i^*}) \geq G(T^*) \geq G(T) - \sum_{i=j+1}^{K-1} p_{iu} r_i + c_u$$

The result follows. □

Note that there are cases where BESTRESERVEBKUP is strictly suboptimal, (e.g., in Figure 1, where OPT probes the backup channel $k$ on some paths). But, in practice, the gain of BESTRESERVEBKUP substantially exceeds the lower bound in Theorem 5.3. For example, even in Figure 1 (which is one of the few cases where we observed the suboptimality of BESTRESERVEBKUP) the gain of RESERVEBKUP($k$), and hence that of BESTRESERVEBKUP, is only $0.1\%$ less than that of OPT.

We now point out an important property of RESERVEBKUP$(0)$. Recall that $\mathcal{P}(0)$ consists of all policies that transmit only in probed channels and never uses backup channels. Thus, RESERVEBKUP$(0)$, which will henceforth be denoted as OPTNOBKUP, attains the maximum gain among all such policies. From Theorem 5.2, OPTNOBKUP can be determined in $O(nK + n\log n)$ time. Thus, the optimum policy is polynomial time computable when backups are not allowed.

Finally, when $K = 2$, every policy is in the class $\cup_{u=0}^{n} P(u)$, and hence Theorem 5.2 also proves that the optimal policy can be computed in $O(n^2)$ time for $K = 2$. However for $K = 2$ we have already obtained an optimal policy which has a lower running time.

## 5.2 Proof of the Structure Theorem

**Definition 5.2.** *A $\leq i$ tree is a decision tree which takes the same decisions irrespective of the states of the probed channels provided the states are less than or equal to $i$. The decisions corresponding to the states which are less than or equal to $i$ therefore constitute a path in such a tree which we refer to as a $\leq i$ path.*

We prove the Structure Theorem 5.1 (in fact a slightly stronger version which we would require later) using the following lemma.

**Lemma 5.5.** *Suppose* OPT *probes a channel $j$ at a node $m$ in its decision tree and if $j$ is in state $i$ it takes a backup downstream. Then there exists another optimum policy* OPT1 *which has the same decision tree as* OPT *except possibly for the tree rooted at $m$. In* OPT1, *the tree rooted at $m$,*

1. *is a $\leq i$ tree*
2. *takes a backup, say $\ell$, at the end of its $\leq i$ path and*

3. *takes $\ell$ as a backup wherever it takes a backup.*

*Proof.* We prove the lemma using induction on the states. The lemma holds by vacuity for all channels $j$, node $m$ if $i = K - 1$. This is because OPT never takes a backup after observing a channel in state $K - 1$.

Suppose the lemma holds for all channels $j$, nodes $m$ in the decision tree of OPT, states $i+1, \ldots, K-1$. We prove the lemma for all channels $j$, nodes $m$ and state $i$. Now, let OPT probe a channel $j$ at a node $m$ in its decision tree and take a backup somewhere downstream if $j$ is in state $i$. Let $m_1$ be the first node after $j$ is probed at $m$ and observed to be in state $i$. Clearly, the decision tree rooted at $m_1$ is a $\leq i$ *tree*, and takes at least one backup somewhere downstream.

We will first show that there is one optimum policy OPT2 which is similar to OPT except possibly for the tree rooted at $m_1$. The tree rooted at $m_1$ in OPT2 is still a $\leq i$ *tree* but

**(p1)** takes a backup, say $\ell$, at the end of its $\leq i$ *path* and

**(p2)** takes $\ell$ as a backup wherever it takes a backup.

Suppose the tree $T$ rooted at $m_1$ in OPT does not satisfy the above conditions. Then there is a path originating from its $\leq i$ *path* which ends in a backup. Let $m_2$ be the highest node (i.e., node closest to $m_1$) from where such a path originates on the $\leq i$ *path* of $T$. Clearly this path corresponds to a channel being observed in a state higher than $i$, say $q$, at $m_2$. From the induction hypothesis, there exists one optimum OPT2 which is similar to OPT except possibly for the decision tree rooted at $m_2$. In OPT2 this decision tree

- is a $\leq q$ *tree*
- takes a backup, say $\ell$, at the end of its $\leq q$ *path* and
- takes $\ell$ as a backup wherever it takes a backup.

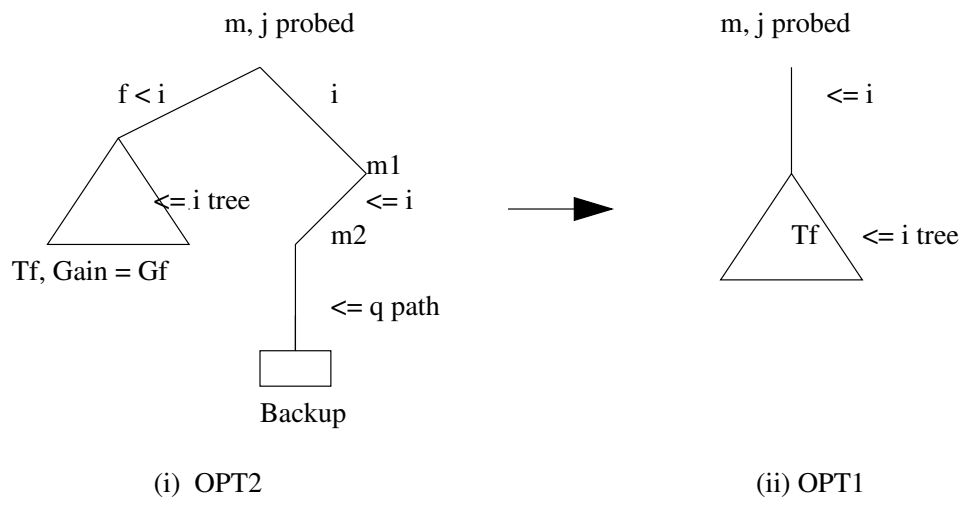

Figure 2: The transformation of OPT2 to OPT1 in lemma 5.5.

Note that OPT2 satisfies conditions **(p1)** and **(p2)** for the tree rooted at $m_1$, see Figure (2).

Let $\alpha$ be the probability that OPT2 never visits $m$, $G'$ be the expected gain of OPT2 if it never visits $m$, $G_h$ be the expected gain of OPT2 given that $j$ is observed in state $h$ at node $m$. Thus, the expected gain of OPT2 is $\alpha G' + (1-\alpha)\sum_h p_{hj} G_h$. Let $T_h$ be the decision tree in OPT2, and hence in OPT, after $j$ is observed to be in a state $h < i$ after being probed at $m$. Consider a new policy which is similar to OPT2 except that it replaces the decision tree rooted at $m_1$ with $T_f$ for some $f < i$. Note that the gain of this new policy given that $j$ is observed in state $i$ at node $m$ is at least $G_f$ since the overall gain after observing a state $i$ and a subsequent sequence of actions can not be less as compared to that after observing $f$ and the same subsequent sequence of actions. Thus, the expected gain of this new policy is at least $\alpha G' + (1-\alpha)\sum_{h, h \neq i} p_{hj} G_h + (1-\alpha) p_{ij} G_f$. Since the expected gain of this policy can not exceed that of the optimum, $G_f \leq G_i$.

Now, consider another policy OPT1 which is similar to OPT2 except that it replaces the decision trees $T_0, \ldots, T_{i-1}$ (i.e., those rooted at nodes immediately downstream of $m$ and corresponding to $j$ being in states lower than $i$), with the decision tree rooted at $m_1$ (i.e., the one corresponding to $j$ being in state $i$), refer Figure (2). Since the decision tree rooted at $m_1$ is a $\leq i$ tree and the $\leq i$ path ends in a backup, the gain of OPT1 is $\alpha G' + (1-\alpha)\left(G_i \sum_{h \leq i} p_{hj} + \sum_{h>i} p_{hj} G_h\right)$. Thus, since $G_f \leq G_i$, for $f < i$, the expected gain of OPT1 is at least as high as that for OPT2. Thus, OPT1 is also optimum. Note that OPT1 is similar to OPT except possibly for the decision tree rooted at $m$, which is a $\leq i$ *tree* and satisfies conditions **(p1)** and **(p2)**. The result follows. □

We now state and prove a theorem which implies the Structure Theorem.

**Theorem 5.6** (Implies the Structure Theorem)**.** *There exists an optimum policy that uses a unique backup channel. If such an optimum policy probes at least one channel, it uses the backup channel at the end of a $\leq i$ path originating from the root of its decision tree.*

*Proof.* Consider an optimum policy that does not use a backup channel at all. Consider the path in its decision tree which corresponds to all probed channels being in state 0. Modify the policy so as to use the last channel probed in this path as a backup. Note that the gain does not decrease. Thus, the modified policy is also optimum. Thus, there always exists an optimum policy that uses a backup channel at the end of some path in its decision tree. If one such optimum policy OPT3 does not probe any channel, then the theorem follows. Let OPT3 probe a channel. Clearly, then, OPT3 probes a channel $j$ at the root node of its decision tree, say $m$. Let $i$ be the highest state of $j$ for which OPT3 transmits in a backup channel somewhere downstream of $m$. Then, by lemma 5.5, there exists another optimum policy OPT4 for which the decision tree rooted at $m$, and hence the overall decision tree, is (a) a $\leq i$ tree (b) uses a channel, say $\ell$, as a backup at the end of the $\leq i$ path in the tree and (c) uses $\ell$ as a backup whenever it uses a backup. The theorem follows. □

### 5.3 The Policy RESERVEBKUP($\ell$): Algorithm and Intuition

We require the subsequent definitions to specify the policy RESERVEBKUP($\ell$).

**Definition 5.3.** *For $i = 1, \ldots, n$,, let $\tilde{r}_i[u] = \frac{\sum_{v=u}^{v=K-1} p_{vi} r_{vi}}{\sum_{v=u}^{v=K-1} p_{vi}}$ and $\tilde{p}_i[u] = \sum_{v=u}^{v=K-1} p_{vi}$. Let $\tilde{r}_0[0] = -1$ and $r_{-1} = -1$. Let $w_\ell = \min_u\{u : r_u > \tilde{r}_\ell[0]\}$*[1].

Note that $\tilde{r}_\ell[0]$ is the probability of success if the sender uses $\ell$ as a backup.

**Definition 5.4.** *Let $H_{u,\ell} = \phi$ for all $u \geq K$. For each $\ell$, starting from $u = K-1$, down to $u = w_\ell$, recursively, define $H_{u,\ell} = \left\{i | i \notin \bigcup_{v:v>u} H_{v,\ell},\ \text{and}\ \tilde{r}_i[u] - \frac{c_i}{\tilde{p}_i[u]} > \max(\tilde{r}_\ell[0], r_{u-1})\right\} \setminus \{\ell\}$. Assume that $c_i/\tilde{p}_i[u] = \infty$ when $\tilde{p}_i[u] = 0$.*

[1] Note that $w_\ell$ is well-defined for each $\ell$ as (a) $\tilde{r}_\ell[0] \geq r_0 = 0$ and (b) $\tilde{r}_\ell[0] < r_{K-1}$ which follows since $p_{K-1\ell} < 1$ for each $\ell$.

RESERVEBKUP($\ell$)

(**Probing Process:**)

$u = K - 1$

While $u \geq w_\ell$ and the highest state of a probed channel is lower than $u$,

probe channels in $H_{u,\ell}$ in non-increasing order of $\tilde{r}_j[u] - c_j/\tilde{p}_j[u]$.

$u \rightarrow u - 1$

(**Selection Process:**)

Consider the channel $j$ in the highest state $y$ among all probed channels.

(If no channel is probed, $j = -1$.)

If $r_y \geq \tilde{r}_\ell[0]$, then transmit in $y$, else transmit in $\ell$.

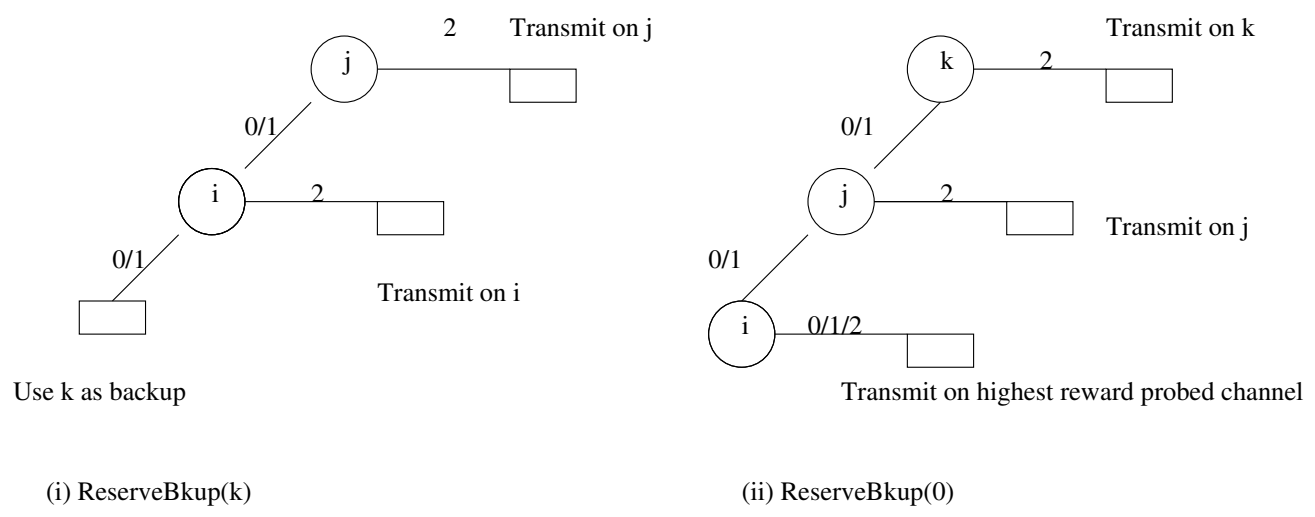

Figure 3: The illustration of RESERVEBKUP($\ell$) in Figure 1.

Refer to Figure (3) for examples elucidating RESERVEBKUP($\ell$). We now explain why RESERVEBKUP($\ell$) attains the same gain as OPT($\ell$) for $\ell = 0, \ldots, n$.

First, observe that RESERVEBKUP($\ell$) $\in \mathcal{P}(\ell)$. This is because by definition $\ell \notin H_{u,\ell}$ for any $u$. Thus, RESERVEBKUP($\ell$) never probes $\ell$ (refer to the probing process). Also, note that RESERVEBKUP($\ell$) does not use any channel other than $\ell$ as backup (refer to the selection process).

Next, since $\tilde{r}_\ell[0]$ is the probability of success when the sender transmits in the backup $\ell$, the channel selection for RESERVEBKUP($\ell$) is clearly optimal among policies that have the same probing sequence.

We now explain the intuition behind the design of the probing process for RESERVEBKUP($\ell$). First let $\ell = 0$. Once a sender observes that a probed channel is in state $u$ it can not increase its gain any further by discovering another probed channel in state $u$ or in a lower state. Thus, subsequently, it probes only the channels $j$ for which the additional reward ($\tilde{r}_j[u+1]\tilde{p}_j[u+1] - r_u\tilde{p}_j[u+1]$) exceeds the cost $c_j$, i.e., it probes the channels in $H_{v,\ell}$, $v > u$. The incremental gain for probing a channel $j$ then is $\tilde{r}_j[u+1]\tilde{p}_j[u+1] - r_u\tilde{p}_j[u+1] - c_j$. The probing sequence in each $H_{u,\ell}$ follows an increasing order of the ratio between this incremental gain and the probability that the channel is in a state that is higher than the highest observed state $u$ ($\tilde{p}_j[u+1]$). We now comment on the major differences between the probing processes of RESERVEBKUP($\ell$) for $\ell > 0$ and $\ell = 0$. Note that $\tilde{r}_\ell[0]$ is the gain if the sender transmits in $\ell$ without probing any channel $j$. If, the sender observes a channel to be in state $u$ for which $r_u \leq \tilde{r}_\ell[0]$, the observation does not increase the gain as compared to $\tilde{r}_\ell[0]$. Hence, the sender considers the incremental gain as $\tilde{r}_j[u+1]\tilde{p}_j[u+1] - \max(r_u, \tilde{r}_\ell[0])\tilde{p}_j[u+1]$ instead of $\tilde{r}_j[u+1]\tilde{p}_j[u+1] - r_u\tilde{p}_j[u+1]$, and, as before, probes only channels for which the incremental gain exceeds the probing cost.

Finally, we determine the computation times of RESERVEBKUP($\ell$) and BESTRESERVEBKUP. All the $H_{u,\ell}$s can be evaluated in $O(nK)$ time and the sorting required to determine the probing sequence within each $H_{u,\ell}$ needs $O(n \log n)$ time. Thus, the entire probing sequence and hence for any given $\ell$ RESERVEBKUP($\ell$), can be computed in $O(nK + nK \log n)$ time or rather in $O(nK\text{log}n)$ time. Now, note that the computations for $H_{u,0}$s and the sorting order in each $H_{u,0}$ can be reused to determine $H_{u,\ell}$s for all $\ell$s in additional $O(n)$ time. Thus, all ReserveBkup($\ell$)s can be computed in $O(nK \log n)$ time. The gain for each ReserveBkup($\ell$) can be computed in $O(nK)$ time. Thus, BestReserveBkup can be computed in $O(n^2K)$ time.

## 5.4 Proof for Theorem 5.2

We now show that the policy ReserveBkup($\ell$) described in Section 5.3 is optimal in the class of policies $\mathcal{P}(\ell)$. This completes the proof for Theorem 5.2 since we have already proved in Section 5.3 that for any given $\ell$ ReserveBkup($\ell$), can be computed in $O(nK \log n)$ time.

First observe that the optimal in the class of policies $\mathcal{P}(\ell)$ need not be unique. We consider Opt($\ell$) to be one optimal policy in $\mathcal{P}(\ell)$ that satisfies the following property. Suppose channel $j$ is the last channel probed in a path in the decision tree of Opt($\ell$); let $m$ be the node at which $j$ is probed. Then, Opt($\ell$) would attain a lesser gain if it were not to probe $j$ at $m$. Clearly, such optimal policies exist, and can be obtained by progressively removing the lowest node at which a channel is probed and which can be removed without reducing the gain.

We first observe the following about Opt($\ell$). Let the highest state of a probed channel be $u$ when Opt($\ell$) terminates its probing process. Then Opt($\ell$) transmits in the probed channel if $r_u \geq \tilde{r}_\ell[0]$ and transmits in $\ell$ otherwise. Thus, the channel selection for ReserveBkup($\ell$) is optimal among all policies that have the same probing sequence. We now state and prove three lemmas which will establish that the probing process for ReserveBkup($\ell$) is optimal in $\mathcal{P}(\ell)$, and thereby prove Theorem 5.2.

**Lemma 5.7.** *1. If all channels in $\bigcup_{v>\max(u,w_\ell-1)} H_{v,\ell}$ have already been probed, and the best state seen so far is u, then Opt($\ell$) does not probe any further.*

*2. Opt($\ell$) can not terminate the probing process when there an is un-probed channel in $\bigcup_{v>\max(u,w_\ell-1)} H_{v,\ell}$, and the best state seen so far is $u$.*

*Proof.* The first part of the lemma clearly holds for $u = K-1$ since the optimal policy does not probe any further after observing a channel in state $K-1$. Let the first part not hold for some $u < K-1$. Thus, although all channels in $\bigcup_{v>\max(u,w_\ell-1)} H_{v,\ell}$ have already been probed, Opt($\ell$) probes further channels. Let $j$ be the last channel probed by Opt($\ell$) in one such path. Let $j$ be probed at node $m$ of the decision tree. The highest state of a channel probed before $j$ is probed at $m$ be $u$. Note that after probing $j$ Opt($\ell$) transmits in (a) backup $\ell$ if the maximum of $u$ and $j$'s state is $w_\ell - 1$ or lower and (b) the channel that has the highest state among all the probed channels otherwise. Now, consider another policy $C \in \mathcal{P}(\ell)$ which is similar to Opt($\ell$) except that it does not probe $j$ at node $m$, and instead transmits in (a) backup $\ell$ if $u < w_\ell$ and (b) the probed channel that is in state $u$ otherwise. Let Opt($\ell$) and hence $C$ reach node $m$ with probability $\alpha$. Clearly, $\alpha > 0$, else node $m$ can be removed from the decision tree of Opt($\ell$) without reducing its gain. Let $\Delta$ be the difference between the gains of Opt($\ell$) and $C$. We will arrive at a contradiction by showing that $\Delta \leq 0$. Hence, the first part of the lemma holds.

$$
\begin{aligned}
\Delta/\alpha &= \sum_{k=0}^{K-1} p_{kj} \max(r_k, r_u, \tilde{r}_\ell[0]) - c_j - \max(r_u, \tilde{r}_\ell[0]) \\
&= \sum_{k=\max(u,w_\ell-1)+1}^{K-1} p_{kj} \left(\max(r_k, r_u, \tilde{r}_\ell[0]) - \max(\tilde{r}_\ell[0], r_u)\right) - c_j \\
&= \sum_{k=\max(u,w_\ell-1)+1}^{K-1} p_{kj} \left(r_k - \max(\tilde{r}_\ell[0], r_u)\right) - c_j.
\end{aligned}
$$

First, let $u \geq w_\ell - 1$. Thus, $j \notin \cup_{v>u} H_{v,\ell}$. Thus, $\Delta < 0$. Now, let $u < w_\ell - 1$. Then, $r_u \leq r_{w_\ell - 1} \leq \tilde{r}_\ell[0]$. Thus, $\Delta/\alpha = \sum_{k=w_\ell}^{K-1} p_{kj}\left(r_k - \max(\tilde{r}_\ell[0], r_{w_\ell - 1})\right) - c_j$. Also, $j \notin \bigcup_{v \geq w_\ell} H_{v,\ell}$. Thus, $\Delta < 0$.

The second part of the lemma holds by vacuity when $u = K-1$ since $H_{K,\ell} = \phi$. Let $u < K-1$. Let the probability that $Opt(\ell)$ terminates the probing process even when there an is un-probed channel in $H_{v,\ell}$ for some $v > \max(u, w_\ell - 1)$, and the best state seen so far is $u$ be $\alpha_1 > 0$. Then, whenever the above event happens, $\text{OPT}(\ell)$ transmits in (a) $\ell$ if $u < w_\ell$ and (b) a probed channel that is in state $u$ otherwise. Consider another policy $D \in \mathcal{P}(\ell)$ which is similar to $\text{OPT}(\ell)$ except that whenever the above event happens it probes an additional channel $j \in H_{v,\ell}$, and transmits in (a) $\ell$ if the maximum of $u$ and $j$'s state is $w_\ell - 1$ or lower and (b) the probed channel that has the highest state otherwise. Let $\Delta_1$ be the difference between the gains of $\text{OPT}(\ell)$ and $D$. We will show that $\Delta_1 < 0$ which is a contradiction. Hence, the second part of the lemma holds.

$$\begin{aligned}
\Delta_1/\alpha_1 &= \max(r_u, \tilde{r}_\ell[0]) - \sum_{k=0}^{K-1} p_{kj} \max(r_k, r_u, \tilde{r}_\ell[0]) + c_j \\
&= c_j - \sum_{k=\max(u, w_\ell - 1)+1}^{K-1} p_{kj}\left(r_k - \max(\tilde{r}_\ell[0], r_u)\right) \\
&= c_j - \sum_{k=\max(u, w_\ell - 1)+1}^{v-1} p_{kj}\left(r_k - \max(\tilde{r}_\ell[0], r_u)\right) - \sum_{k=v}^{K-1} p_{kj}\left(r_k - \max(\tilde{r}_\ell[0], r_u)\right) \\
&\leq c_j - \sum_{k=v}^{K-1} p_{kj}\left(r_k - \max(\tilde{r}_\ell[0], r_{v-1})\right).
\end{aligned}$$

The last inequality follows since $r_k \geq \max(\tilde{r}_\ell[0], r_u)$ for $k > \max(u, w_\ell - 1)$ and $r_{v-1} \geq r_u$ since $v \geq \max(u, w_\ell - 1) + 1$. The result follows since $j \in H_{v,\ell}$. $\square$

**Lemma 5.8.** *Let $w_\ell \leq u < K$. $\text{OPT}(\ell)$ probes all channels in $H_{u,\ell}$ before probing any channel that is not in $\cup_{v \geq u} H_{v,\ell}$. Also, $\text{OPT}(\ell)$ probes all channels of $H_{u,\ell}$ unless one of the probed channels is in state $u$ or a higher state, and probes these channels in non-increasing order of $\tilde{r}_j[u] - \frac{c_j}{\tilde{p}_j[u]}$.*

*Proof.* Suppose the lemma does not hold. Then there exists a node in the decision tree of $\text{OPT}(\ell)$, which $\text{OPT}(\ell)$ visits with positive probability[2], such that the decisions at it violate the lemma. Let $m$ be such a node which is also the farthest from the root node among those that violate this lemma. Then there exists a state $q \geq w_\ell$ such that there exists a channel in $H_{q,\ell}$ that has not been probed by $\text{OPT}(\ell)$ before it visits $m$ and the best state seen so far is $q-1$ or worse. Let $u$ be the highest state that satisfies both the above criteria and let $j$ be the channel with the largest $\tilde{r}_j[u] - \frac{c_j}{\tilde{p}_j[u]}$ value among the un-probed channels of $H_{u,\ell}$. At node $m$, $\text{OPT}(\ell)$ does not probe $j$ contradicting the lemma. From the second part of lemma 5.7, the probing process of $\text{OPT}(\ell)$ can not terminate at $m$. Thus, $\text{OPT}(\ell)$ probes some channel $i \neq j$ at node $m$. Note that $i \notin \cup_{v>u} H_{v,\ell}$.

Since $\text{OPT}(\ell)$ have already probed all channels in $\cup_{v>u} H_{u,\ell}$, if channel $i$ is in state $u$ or a higher state, $\text{OPT}(\ell)$ does not probe any further (first part of lemma 5.7), and transmits in $i$ (since $i$ has the highest state, say $s$, among the probed channels and $r_s \geq r_u \geq r_{w_\ell} > \tilde{r}_\ell[0]$). If $i$ is in a state $u-1$ or a lower state, since $H_{u,\ell}$ has un-probed channels, the probing process can not terminate (second part of lemma 5.7). Now, $\text{OPT}(\ell)$ probes $j$ next (otherwise $m$ will not be the node farthest from the root to violate the lemma). Using similar arguments, it follows that after probing $j$, $\text{OPT}(\ell)$ transmits in $j$ if $j$ is in state $u$ or a higher state.

[2] If the lemma is violated at a node which $\text{OPT}(\ell)$ visits with 0 probability, we can, without reducing the gain of $\text{OPT}(\ell)$, alter the decisions at the node so as to satisfy the lemma. Hence, without any loss of generality, we assume that the decisions of $\text{OPT}(\ell)$ satisfy the lemma at all such nodes.

The situation resembles the decision tree in Figure (4a). The trees $T_1 \dots T_{u^2}$ correspond to observing the ordered pair $(i = u', j = u'')$ where $0 \leq u', u'' \leq u - 1$. The square boxes denote that $\text{OPT}(\ell)$ do not probe anything else.

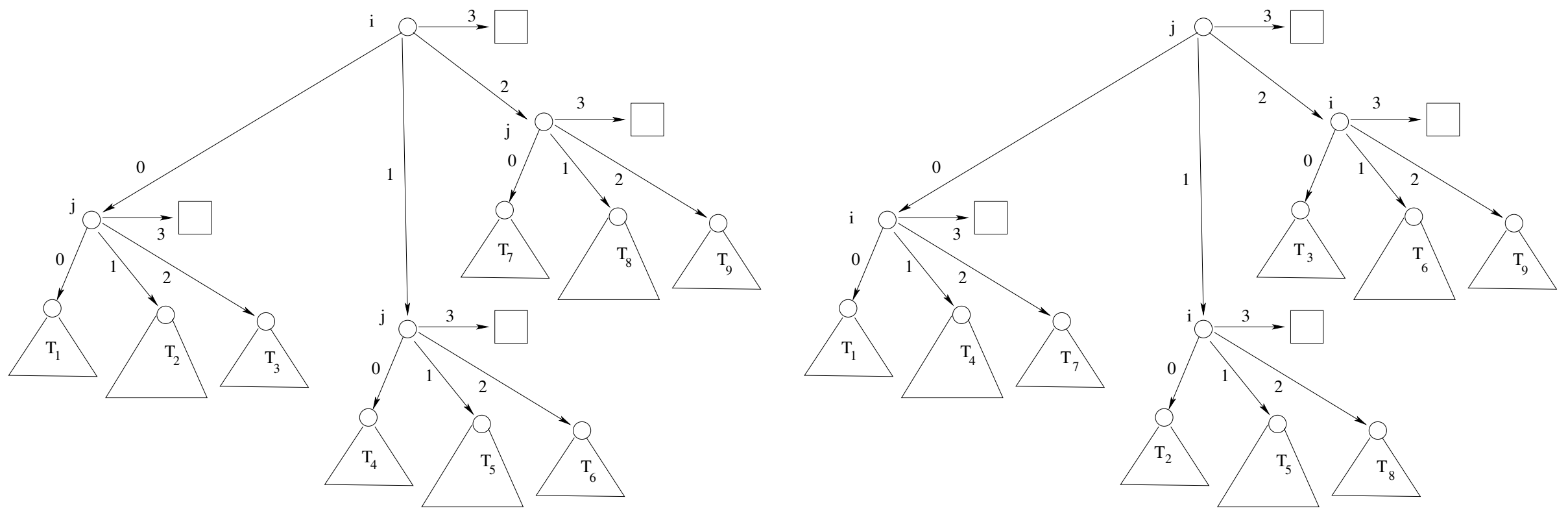

(a) (b)

Figure 4: The decision trees of OPT for $u = 3$

Let $\text{OPT}(\ell)$ not traverse node $m$ with probability $\alpha_1$, traverse node $m$ and stop after probing $i$ or $j$ with probability $\alpha_2$, traverse node $m$ and continue after probing $j$ with probability $\alpha_3$. By assumption, $\alpha_2 > 0$. Let the conditional expected gains given these scenarios be $G_1, G_2, G_3$ respectively. Then, the expected gain of $\text{OPT}(\ell)$ is $G_{\text{Opt}(\ell)} = \sum_{i=1}^{3} \alpha_i G_i$. Let the total probing cost en-route to node $m$ be $C_1$. Then, $G_2 = \frac{1-\alpha_1}{\alpha_2} \left(\tilde{p}_i[u]\tilde{r}_i[u] - c_i + (1 - \tilde{p}_i[u])[\tilde{p}_j[u]\tilde{r}_j[u] - c_j] - C_1\right)$.

Now consider an alternate policy $A$ that is similar to $\text{OPT}(\ell)$ except for the tree rooted at node $m$. Figure (4) shows the tree rooted at node $m$ in policy $A$ for $u = 3$. Here, $A$ probes $j$ at node $m$ and subsequently probes $i$ unless $j$ is in state $u$ or a higher state. The tree $T'$ corresponding to the ordered pair $(i = u', j = u'')$ in OPT is assigned on the branch corresponding to the ordered pair $(j = u'', i = u')$ in $A$. The contributions to the gain from the trees $T_1, \dots T_{u^2}$ remain the same because in both the scenarios the probabilities of probing these trees are the same. Thus, the expected gain of this new policy is $G_C = \alpha_1 G_1 + \alpha_2 G_2' + \alpha_3 G_3$, where $G_2' = \frac{1-\alpha_1}{\alpha_2} \left(\tilde{p}_j[u]\tilde{r}_j[u] - c_j + (1 - \tilde{p}_j[u])[\tilde{p}_i[u]\tilde{r}_i[u] - c_i] - C_1\right)$.

Now *if* $i \in H_{u,\ell}$ then we have $\tilde{r}_j[u] - \frac{c_j}{\tilde{p}_j[u]} > \tilde{r}_i[u] - \frac{c_i}{\tilde{p}_j[u]}$ which is the condition that arises from violating the non-increasing order. If $i \notin H_{u,\ell}$, then since $i \notin \cup_{v>u} H_{v,\ell}$ and $u \geq w_\ell$, $\tilde{r}_i[u] - \frac{c_i}{\tilde{p}_i[u]} \leq \max(r_{u-1}, \tilde{r}_\ell[0])$. But $\tilde{r}_j[u] - \frac{c_j}{\tilde{p}_j[u]} > \max(r_{u-1}, \tilde{r}_\ell[0])$ since $j \in H_{u,\ell}$ and $u \geq w_\ell$. Therefore, *in both cases* we have $\tilde{r}_j[u] - \frac{c_j}{\tilde{p}_j[u]} > \tilde{r}_i[u] - \frac{c_i}{\tilde{p}_i[u]}$. But this implies that

$$
\begin{aligned}
& \frac{G_C - G_{\text{Opt}(\ell)}}{1 - \alpha_1} \\
= \; & \tilde{p}_j[u]\tilde{r}_j[u] - c_j + (1 - \tilde{p}_j[u]) \{\tilde{p}_i[u]\tilde{r}_i[u] - c_i\} \\
& -\tilde{p}_i[u]\tilde{r}_i[u] + c_i - (1 - \tilde{p}_i[u]) \{\tilde{p}_j[u]\tilde{r}_j[u] - c_j\} \\
= \; & \tilde{p}_i[u]\tilde{p}_j[u] \left( \tilde{r}_j[u] - \frac{c_j}{\tilde{p}_j[u]} - \tilde{r}_i[u] + \frac{c_i}{\tilde{p}_i[u]} \right) > 0.
\end{aligned}
$$

Thus, since $\alpha_1 < 1$, $G_C > G_{\text{Opt}(\ell)}$. Thus, we arrive at a contradiction. The result follows. □

**Lemma 5.9.** $\text{OPT}(\ell)$ *probes only channels in* $\cup_{v=w_\ell}^{K-1} H_{v,\ell}$.

*Proof.* From the first part of lemma 5.7, OPT($\ell$) terminates its probing process after probing all channels in $\cup_{v=w_\ell}^{K-1} H_{v,\ell}$. From lemma 5.8, OPT($\ell$) must probe all channels in $\cup_{v=w_\ell}^{K-1} H_{v,\ell}$ before probing a channel that is not in $\cup_{v=w_\ell}^{K-1} H_{v,\ell}$. The result follows. □

The optimality of the probing process for RESERVEBKUP($\ell$) in $\mathcal{P}(\ell)$ follows from lemmas 5.8 and 5.9. Thus, Theorem 5.2 follows.

# 6 Additive Approximation Schemes for Equal Probing Costs

We now consider the case that all channels have equal probing costs, i.e., $c_i = c > 0$, but still allow for different distributions for the state processes of different channels. This assumption is motivated by the fact that oftentimes the probing cost is determined by the energy consumed in transmitting the probe packets which is usually similar for different channels. We still assume that the sender is saturated. We present a policy that given any $\epsilon > 0$ attains a gain of at least $G_{\text{OPT}} - \epsilon r_{K-1}$. The time to compute this policy depends exponentially on $\frac{1}{\epsilon}$, but is a polynomial in $n$ for any fixed $\epsilon > 0$.

Motivated by the Structure Theorem (Theorem 5.6), we consider the following class of policies.

**Definition 6.1.** *Let $P(\ell, i)$ be the class of policies which (a) take the same actions if a probed channel is observed in state $i'$ or $i''$ at any node when $i', i'' \leq i$ and (b) take backup $\ell$ at the end of the $\leq i$ path originating from the roots of their decision trees and do not take backups anywhere else. Let Opt($\ell, i$) denote the optimum policy in this class with gain $G(\ell, i)$.*

We know from Theorem 5.6 that OPT is in $P(\ell^*, i^*)$ for some $\ell^* \in \{1, \ldots, n\}$ and some $i^* \in \{0, \ldots, K-1\}$. Therefore it suffices to provide approximations of the *Opt*($\ell, i$) policies for different $\ell, i$. Note that the policy that approximates *Opt*($\ell, i$) need not be in $P(\ell, i)$.

We now prove the central lemma in this section, which also presents the approximation algorithm.

**Lemma 6.1.** *We can compute in $O(n^{h+2}hK)$ time a policy whose gain is at least $G(\ell, i) - \epsilon r_{k-1}$, where $h = 1 + \lceil -\log_{1/(1-\epsilon)} \epsilon \rceil$.*

*Proof.* First, let $c \leq \epsilon r_{K-1}$. Now, observe that OPTNOBKUP attains a gain of at least $G(\ell, i) - \epsilon r_{K-1}$ since it attains a gain of at least $G(\ell, i) - c$. To see the latter, note that if *Opt*($\ell, i$) is modified to first probe $\ell$ and subsequently transmit in $\ell$ wherever it were using $\ell$ as a backup, then its gain decreases by at most $c$. Thus, the modified policy has a gain of at least $G(\ell, i) - c$. Now, since the modified policy does not use any backup channel, its gain is at most that of OPTNOBKUP. The result follows.

Now, let $c > \epsilon r_{K-1}$. Note that *Opt*($\ell, i$) does not transmit in a probed channel whose state is $i$ or less. Now, using a proof which is the same as that for Claim 5.4, it follows that if *Opt*($\ell, i$) probes a channel $u \neq \ell$, and if $u$ is not the first channel probed, then $\sum_{j=i+1}^{K-1} p_{ju} r_j \geq c > \epsilon r_{K-1}$, and hence $\sum_{j=i+1}^{K-1} p_{ju} > \epsilon$. Now, consider the $\leq i-$path of *Opt*($\ell, i$) starting from the root, which ends in the backup. The probability of continuing on this path decreases by a factor of $1 - \epsilon$ for each additional node after the first node (escape probability is at least $\epsilon$ after the first node). Therefore, the probability $q$ of continuing beyond $h$ nodes in this path is less than $\epsilon$, since $h = 1 + \lceil -\log_{1/(1-\epsilon)} \epsilon \rceil$. Thus, if *Opt*($\ell, i$) is modified to take the backup after $h$ nodes in this path, the gain of the modified policy is at least $G(l, i) - \epsilon r_{K-1}$. Let $P(\ell, i, h)$ be the class of policies that are in $P(\ell, i)$ and have $h$ or fewer nodes in their $\leq i$ path originating from their roots. Thus, the policy that has the maximum gain among all policies in $P(\ell, i, h)$ has gain at least $G(l, i) - \epsilon r_{K-1}$.

Thus, the result follows if we can compute the policy that has the maximum gain in $P(\ell, i, h)$ in $O(n^{h+2}hK)$ time. Note that a channel can be probed at most once in the $\leq i$ path originating from the root for this policy, and thus the number of possible probing sequences for this path is bounded by $O(n^h)$. For each probing sequence in this path, we compute as follows the rest of the actions so as to maximize the gain. Suppose that we are at a node $t$ in this path where we just probed channel $x$ and the set of probed channels (including $x$) is $\mathcal{Q}$. Since the probing sequence in the $\leq i$ path is given, we only need to determine the actions downstream if $x$ is in state $j > i$. In this case, we know from Definition 6.1, that a backup is not used downstream. Also, only the channels in $\overline{\mathcal{Q}}$ can be probed downstream. Further, all channels which are in states $\leq j$ will not be used for transmission, and probing them do not increase the gain. Therefore we can pretend that we have a new system over $\overline{\mathcal{Q}}$ where $r''_s = r_s - r_j$ if $s > j$ and 0 otherwise. We can use OPTNOBKUP on the channels of $\overline{\mathcal{Q}}$ with rewards $\{r''\}$ to find an optimal subtree.

Thus, given the probing sequence in the $\leq i$ path, the rest of the tree can be computed using $h$ applications of OPTNOBKUP, which requires $O(hnK \log n))$ time. The gain of each tree can be computed in $O(nKh)$ time. Thus, $O(hnK \log n)$ time is needed in this part. Since there are $O(n^h)$ probing sequences for the $\leq i$ path, the policy that attains the maximum gain in $P(\ell, i, h)$ can be computed in $O(n^{h+2}hK)$ time. The result follows. □

From lemma 6.1 and since $G_{\text{OPT}} = G(\ell, i)$, for some $\ell \in \{1, \dots, n\}$ and $i \in \{0, \dots, K-1\}$, the policy that has the maximum gain among those that attain gains of $G(\ell, i) - \epsilon r_{K-1}$ for different $\ell \in \{1, \dots, n\}$ and $i \in \{0, \dots, K-1\}$ attains a gain of at least $G_{\text{OPT}} - \epsilon r_{K-1}$. We can compute such a policy in $O(n^{h+3}hK^2)$ time. We therefore obtain the following theorem.

**Theorem 6.2.** *In $O(n^{h+3}hK^2)$ time we can compute a policy whose gain is at least $G_{\text{OPT}} - \epsilon r_{K-1}$ where $h = 1 + \lceil -\log_{1/(1-\epsilon)} \epsilon \rceil$.*

Finally, note that since we are focusing on only an additive approximation, the computation time can be made independent of $K$. First, divide $[0, r_{K-1}]$ in disjoint intervals of size $\epsilon/2$. Then, consider a new system where the probability of success in each state $i$ equals $(\epsilon/2)\lfloor 2r_i/\epsilon \rfloor$. This new system effectively consists of at most $2r_{K-1}/\epsilon \leq 2/\epsilon$ states. In this system, the approximate policy we just developed, approximates the optimum gain within an additive factor of $\epsilon/2$. The gain of the optimum policy in this system is at least $G_{\text{Opt}} - \epsilon/2$. Thus, the approximate policy computed in this system attains a gain of at least $G_{\text{Opt}} - \epsilon$ in the actual system. Note that irrespective of the number of states in the original system the time required for computing the approximate policy in the new system is $O(n^{h+3}h/\epsilon^2)$ where $h = 1 + \lceil -\log_{1/(1-\epsilon/2)} \epsilon/2 \rceil$.

# 7 The Unsaturated Sender Problem

We now consider the case that the sender may not always have packets to transmit. We assume that the sender generates packets as per a positive recurrent Markovian arrival process such that the average number of packets arriving in its queue under the steady state distribution of the arrival process is $\lambda$. Thus, the sender may choose not to transmit in a given slot even when her queue is non-empty, e.g., when the transmission conditions of the probed channels are not acceptable for her. However, the sender needs to transmit at rate $\lambda$, else her queue will become unstable. Thus, we need to jointly optimize the probing, channel selection and transmission decisions so as to maximize the gain subject to stabilizing the sender's queue. Specifically, we seek to solve **Problem** 2 formulated in Section 3. As stated in Section 3, we consider $n$ channels with $K$ states and potentially unequal probing costs and different distributions for the state processes. We note that no previous results – even exponential time policies, were known for this problem.

We will assume that the optimal policy OPTUNSAT is ergodic, and denote its gain by $G_U$. We present a stable policy that (a) attains a gain of $G_U$ for $K = 2$ and at least $(2/3)G_U$ for $K > 2$ and (b) can be computed in $O(n^2K(n + K))$ time.

## 7.1 Roadmap and Main Results

We first introduce the following definitions.

**Definition 7.1.** *Let $\Pi$ denote the set of decision trees. Let $C_\sigma$ denote the expected probing cost in decision tree $\sigma \in \Pi$, and $\mathcal{M}_\sigma$ be the set of leaf nodes where the decision is to transmit. Let $\hat{p}_{m\sigma}$ denote the probability that the leaf node $m$ is reached in $\sigma$, and if $m \in \mathcal{M}_\sigma$, $\hat{r}_{m\sigma}$ is the probability of success for the transmission at $m$. Let $\mathcal{S}_\sigma = \sum_{m \in \mathcal{M}_\sigma} \hat{p}_{m\sigma}$ and $\mathcal{G}_\sigma = \sum_{m \in \mathcal{M}_\sigma} \hat{r}_{m\sigma}\hat{p}_{m\sigma} - C_\sigma$ for a $\sigma \in \Pi$.*

Since the number of channels and the number of transmission states of the channels are both finite, $\Pi$ constitutes a finite set. Note that now a decision tree may not transmit in all leaf nodes. For example, the decision trees in Figure 1(a),(b) are examples of decision trees in $\Pi$, and in addition, if the decision trees in Figure 1(b) are modified so as not to transmit at the end of the left-most paths, the modifications will also constitute decision trees in $\Pi$. Note that $\Pi$ also includes the decision tree that neither probes nor transmits in any channel. Thus, if the sender takes actions as per the decision tree $\sigma$ in a slot in which she has packets to transmit, she transmits with probability $\mathcal{S}_\sigma$ in the slot and attains a gain of $\mathcal{G}_\sigma$ in the slot.

**Step 1: Expressing the optimal policy as the solution of a Linear Program:** Throughout this discussion, we assume that $\epsilon \in (0, \frac{1}{\lambda} - 1)$ is a suitably chosen small constant. Consider the following linear program LPUNSAT($\epsilon$).

$$\text{LPUNSAT}(\epsilon): \quad \text{Maximize} \sum_{\sigma \in \Pi} \beta_\sigma \mathcal{G}_\sigma$$

$$\begin{aligned} \textstyle\sum_{\sigma \in \Pi} \beta_\sigma \mathcal{S}_\sigma &= \lambda(1+\epsilon) \qquad (\textit{stability constraint}) \\ \textstyle\sum_{\sigma \in \Pi} \beta_\sigma &= 1 \\ \beta_\sigma &\geq 0 \ \forall \ \sigma \in \Pi \end{aligned}$$

**Definition 7.2.** *Let $\{\beta^*(\epsilon)\}$ be the optimum solution of* LPUNSAT($\epsilon$). *Let $Q^*(\epsilon)$ denote the optimal value of the objective function.*

We will prove that an arbitrary close approximation for the optimal policy can be obtained using $\{\beta^*(\epsilon)\}$. But, $\{\beta^*(\epsilon)\}$ does not provide an exact solution because the stability constraint in LPUNSAT($\epsilon$) involves a positive $\epsilon$, which is required to ensure stability. Thus, the approximation improves with decrease of $\epsilon$. We first prove the following results.

**Proposition 7.1.** *For any $0 \leq \epsilon_1 < \epsilon_2 < \frac{1}{\lambda} - 1$, $Q^*(\epsilon_1) \leq Q^*(\epsilon_2)$.*

*Proof.* Let $\{\beta\}$ denote the optimal solution for LPUNSAT($\epsilon_1$). Thus, $\sum_\sigma \beta_\sigma \mathcal{S}_\sigma = \lambda(1+\epsilon_1) < \lambda(1+\epsilon_2)$.There exists a policy $x$ such that $\beta_x > 0$ and $\mathcal{S}_x < 1$, else $\sum_\sigma \beta_\sigma \mathcal{S}_\sigma = 1 > \lambda(1+\epsilon_1)$. Let $x^C$ denote the policy which transmits at all leaf nodes, but is otherwise similar to $x$. Thus, $\mathcal{S}_{x^C} = 1$, and $\mathcal{G}_{x^C} \geq \mathcal{G}_x$. Now increase $\beta_{x^C}$ and decrease $\beta_x$ such that their sum remains the same. This change increases $\sum_\sigma \beta_\sigma \mathcal{S}_\sigma$, ensures that the objective value does not decrease and that the $\sum_\sigma \beta_\sigma$ remains the same. Continue this process until $\sum_\sigma \beta_\sigma \mathcal{S}_\sigma = \lambda(1+\epsilon_2)$. This yields a feasible solution to LPUNSAT($\epsilon_2$) whose value is at least $Q^*(\epsilon_1)$. □

The next lemma provides an upper bound for the gain of the optimal policy.

**Lemma 7.1.** $G_U \leq Q^*(\epsilon) \ \forall \ \epsilon \in [0, \frac{1}{\lambda} - 1)$.

*Proof.* Let $\beta_\sigma^U$ denote the probability with which the optimal policy OPTUNSAT chooses policy $\sigma$. Then $\{\beta_\sigma^U\}$ forms a feasible solution to LPUNSAT(0), and the expected gain of this policy is simply $\sum_{\sigma \in \Pi} \beta_\sigma^U \mathcal{G}_\sigma$. Thus, $\mathcal{G}_U \leq Q^*(0)$. By Proposition 7.1, we have $Q^*(0) \leq Q^*(\epsilon)$, which completes the proof. □

We next show that any feasible solution $\{\beta(\epsilon)\}$ to LPUNSAT($\epsilon$) yields a stable policy, UNSAT($\beta(\epsilon)$), whose gain is close to the corresponding objective value of LPUNSAT($\epsilon$). We describe policy UNSAT($\beta(\epsilon)$) after introducing the following definition. A slot in which the sender's queue is non-empty is referred to as a *busy* slot.

| Policy UNSAT($\beta(\epsilon)$) |
|---|
| In each busy slot, select a $\sigma \in \Pi$ in accordance with the probability distribution $\{\beta(\epsilon)\}$, and probe channels, decide whether to transmit, and select a channel as per $\sigma$. |

**Lemma 7.2.** *Assume that* $\epsilon \in (0, \frac{1}{\lambda} - 1)$. *If* $\{\beta(\epsilon)\}$ *is a feasible solution of* LPUNSAT($\epsilon$) *and has objective value* $Q(\epsilon)$*, then* UNSAT($\beta(\epsilon)$) *is stable and attains a gain of* $\frac{Q(\epsilon)}{1+\epsilon}$.

*Proof.* In any busy slot, UNSAT($\beta(\epsilon)$) selects a decision tree $\sigma$ in accordance with the probability distribution $\beta(\epsilon)$. Thus, the stability constraint in LPUNSAT($\epsilon$) ensures that in any busy slot the sender transmits packets with probability $\lambda(1+\epsilon)$ which exceeds $\lambda$. Since the state of the arrival process and the queue length under UNSAT($\beta(\epsilon)$) constitutes a Markov chain, stability follows from standard results and analytical techniques (Theorem 2.2.3 in [12], [9]).

Since the system is stable and UNSAT($\beta(\epsilon)$) transmits a packet with probability $\lambda(1+\epsilon)$ in each busy slot, using Little's law, at least $\frac{1}{1+\epsilon}$ of slots are busy. The gain of UNSAT($\beta(\epsilon)$) in each busy slot is $Q(\epsilon)$. Thus, UNSAT($\beta(\epsilon)$) attains a gain of at least $Q(\epsilon)/(1+\epsilon)$. ☐

In view of Lemmas 7.1 and 7.2, UNSAT($\beta^*(\epsilon)$) *is stable and attains a gain of* $\frac{Q^*(\epsilon)}{1+\epsilon} \geq G_U/(1+\epsilon)$. But, we do not know how to solve LPUNSAT($\epsilon$) in polynomial time. We therefore seek to obtain constant factor approximations of the optimal solution of LPUNSAT($\epsilon$) in polynomial time. This motivates the following definitions.

**Definition 7.3.** *A* $c$*-approximation to* LPUNSAT($\epsilon$) *is a feasible solution* $\{\beta\}$ *of* LPUNSAT($\epsilon$) *for which the objective function is at least* $cQ^*(\epsilon)$. *A* $c$*-approximation to the unsaturated sender problem constitutes a (potentially randomized) stable policy that attains a gain of at least* $cG_U$.

It follows from Lemmas 7.1 and 7.2 that for any $\epsilon \in (0, \frac{1}{\lambda} - 1)$, a $c$-approximation to LPUNSAT($\epsilon$) easily yields a $c/(1+\epsilon)$-approximation to the unsaturated sender problem. We therefore focus on obtaining a $c$-approximation to LPUNSAT($\epsilon$) in polynomial time.

**Step 2: Considering the Lagrangean Relaxation.** We consider a Lagrangean Relaxation LPLAGRANGE($\epsilon, \mathcal{L}$) for $\mathcal{L} \geq 0$.

$$\text{LPLAGRANGE}(\epsilon, \mathcal{L}): \quad \text{Maximize} \sum_{\sigma \in \Pi} \beta_\sigma \mathcal{G}_\sigma + \mathcal{L}\left(\lambda(1+\epsilon) - \sum_{\sigma \in \Pi} \beta_\sigma \mathcal{S}_\sigma\right)$$

$$\begin{aligned} \textstyle\sum_{\sigma \in \Pi} \beta_\sigma &= 1 \\ \beta_\sigma &\geq 0 \ \forall \ \sigma \in \Pi \end{aligned}$$

Note that the optimal solution of LPLAGRANGE($\epsilon, \mathcal{L}$) uses only (a) the $\sigma$s that always transmit when $\mathcal{L} \leq 0$ and (b) the $\sigma$s that never transmit when $\mathcal{L} > r_{K-1}$. The hope is that the ideal Lagrange multiplier $\mathcal{L}^*$ would ensure that $\sum_{\sigma \in \Pi} \beta_\sigma \mathcal{S}_\sigma = \lambda(1+\epsilon)$ for the optimum solution of LPLAGRANGE($\epsilon, \mathcal{L}^*$), and we would have a solution for LPUNSAT($\epsilon$). However, the computation time for finding such a $\mathcal{L}^*$ is the same as that for the original problem!

We proceed as follows to circumvent this difficulty. We obtain a $c-$approximate solution for LPUNSAT($\epsilon$) in polynomial time using the following observation and the subsequent lemma.

**Proposition 7.2.** *For any $\mathcal{L} \geq 0$, there exists an optimum solution of* LPLAGRANGE$(\epsilon, \mathcal{L})$ *in which $\beta_\sigma = 1$ for some $\sigma = \sigma^{\mathcal{L}}$, and $\beta_\sigma = 0$ if $\sigma \in \Pi \setminus \{\sigma^{\mathcal{L}}\}$.*

The above proposition motivates the following definition.

**Definition 7.4.** *For any $\mathcal{L} \geq 0$, a policy $\sigma \in \Pi$ is said to c-approximate* LPLAGRANGE$(\epsilon, \mathcal{L})$ *if $\mathcal{G}_\sigma - \mathcal{L}\mathcal{S}_\sigma \geq c(\mathcal{G}_{\sigma'} - \mathcal{L}\mathcal{S}'_\sigma)$ for any $\sigma' \in \Pi$.*

**Lemma 7.3.** *Assume that $\epsilon \in (0, \frac{1}{\lambda} - 1)$ and $0 \leq c \leq 1$. Suppose we have two decision trees $\sigma_+, \sigma_-$ that c-approximate* LPLAGRANGE$(\epsilon, \mathcal{L}^+)$ *and* LPLAGRANGE$(\epsilon, \mathcal{L}^-)$ *respectively. Suppose, further that $\mathcal{S}_{\sigma_+} \leq \lambda(1+\epsilon) < \mathcal{S}_{\sigma_-}$ and that $0 \leq \mathcal{L}^+ - \mathcal{L}^- \leq c\epsilon Q^*(\epsilon)$. Consider $\{\beta\}$ such that $\beta_{\sigma^+} = \alpha$, $\beta_{\sigma^-} = 1 - \alpha$, and $\beta_\sigma = 0$ for $\sigma \in \Pi \setminus \{\sigma_+, \sigma_-\}$ where $\alpha = \frac{\mathcal{S}_{\sigma_-} - \lambda(1+\epsilon)}{\mathcal{S}_{\sigma_-} - \mathcal{S}_{\sigma_+}}$.*

*Then $\{\beta\}$ constitutes a feasible solution for* LPUNSAT$(\epsilon)$ *and $\sum_{\sigma \in \{\sigma_{\mathcal{L}^+}, \sigma_{\mathcal{L}^-}\}} \beta_\sigma \mathcal{G}_\sigma \geq c(1-\epsilon)Q^*(\epsilon)$.*

We prove the above lemma in Section 7.2. Lemmas 7.1, 7.2, 7.3 imply the following fact.

**Proposition 7.3.** *If $\{\beta\}$ satisfies the conditions in lemma 7.3,* UNSAT$(\beta)$ *is a $c(1-\epsilon)/(1+\epsilon)$-approximation to the unsaturated sender problem.*

**Step 3: Finding the two solutions.** We now address the following important issues: (1) how to obtain $c$-approximations for LPLAGRANGE$(\epsilon, \mathcal{L})$ for arbitrary $\mathcal{L}$ and (2) how to obtain two $\mathcal{L}^+, \mathcal{L}^-$ such that the respective $c$-approximations $\sigma_+, \sigma_-$ for LPLAGRANGE$(\epsilon, \mathcal{L}^+)$, LPLAGRANGE$(\epsilon, \mathcal{L}^-)$ satisfy $\mathcal{S}_{\sigma_+} \leq \lambda(1+\epsilon) < \mathcal{S}_{\sigma_-}$. We first observe that the objective function of LPLAGRANGE$(\epsilon, \mathcal{L})$ can be expressed as

$$\sum_{\sigma \in \Pi} \beta_\sigma \mathcal{G}_\sigma + \mathcal{L}\left(\lambda(1+\epsilon) - \sum_{\sigma \in \Pi} \beta_\sigma \mathcal{S}_\sigma\right) = \mathcal{L}(\lambda(1+\epsilon)) + \sum_{\sigma \in \Pi} \beta_\sigma \left(\sum_{m \in \mathcal{M}_\sigma} (\hat{r}_{m\sigma} - \mathcal{L})\hat{p}_{m\sigma} - C_\sigma\right).$$

Thus, optimizing or approximating the above quantity is similar to optimizing or approximating the saturated sender problem in a system where (a) the reward of transmitting in a channel in state $m$ is $r'_m = r_m - \mathcal{L}$ and (b) the sender may choose not to transmit in a slot. The shift in the rewards and the option of not transmitting leads to some important differences with the saturated sender problem we considered earlier (Problem 1). Specifically, the optimal policy in this system will not transmit in a probed channel that is in a state $m$ such that $r_m < \mathcal{L}$, but may transmit in a backup channel in such a state $m$. Thus, the reward of transmitting in a probed channel is non-negative, whereas the reward of transmitting in a backup channel may be negative. Owing to these differences, the proof of the $4/5$-approximation no longer holds in this system. Nevertheless, we obtain a $2/3$-approximation for this system. We first introduce the following definitions.

**Definition 7.5.** *Consider a system where the sender (a) is saturated (i.e., always has packets to transmit) (b) attains a reward of $r_m - x$ if it transmits in a channel in state $m$ (c) acquires a cost of $c_i$ when it probes channel $i$ and (d) may choose not to transmit in a slot. We refer to this system as the* SATURATED ALTERED REWARD $(x)$ *system, and let $T(x)$ be the problem of maximizing the gain in this system. A policy is said to c-approximate the $T(x)$ problem if its gain in this system is at least c times the maximum gain in this system.*

*Note that the policy that solves (c-approximates, respectively) the $T(\mathcal{L})$ problem solves (c-approximates, respectively) the* LPLAGRANGE$(\epsilon, \mathcal{L})$ *problem.*

**Lemma 7.4.** *We can solve $T(x)$ optimally ($c = 1$) for $K = 2$ and achieve a $c = 2/3$ approximation for $K > 2$ in $O(n^2K)$ time.*

The optimum policy in a class of 'threshold type" policies $\textsc{ReserveBkup}(\ell, x)$ provides the above optimal and approximate solutions for different values of $K$; as the name suggests, these threshold-type policies are extensions of $\textsc{ReserveBkup}(\ell)$. We present these threshold type policies in Section 7.3, and prove lemma 7.4 using these policies in Section 7.4.

We finally prove in Section 7.5 the last piece, namely:

**Lemma 7.5.** *Assume that $\epsilon \in (0, \frac{1}{\lambda} - 1)$. Using $O\left(n^2 K(n+K)\right)$ time, we can compute $\sigma_+, \sigma_-, \mathcal{L}^+, \mathcal{L}^-$ which satisfy the properties of lemma 7.3 for (a) $c = 1$ for $K = 2$ and (b) $c = 2/3$ for $K > 2$.*

We present a constructive policy for computing the above $\sigma_+, \sigma_-, \mathcal{L}^+, \mathcal{L}^-$ as part of our proof.

The following Theorem follows from proposition 7.3 and lemma 7.5.

**Theorem 7.6.** *Assume that $\epsilon \in (0, \frac{1}{\lambda} - 1)$. We can compute a $c(1-\epsilon)/(1+\epsilon)$-approximation for the unsaturated sender problem where $c = 1$ for $K = 2$ and $c = 2/3$ for $K > 2$ in $O\left(n^2 K(n+K)\right)$ time.*

Since the computation time does not depend on $\epsilon$, by selecting small $\epsilon$, we can attain in polynomial time an approximation factor close to 1 for $K = 2$ and $2/3$ for $K > 2$. We summarize the policy that attains the above performance guarantee in Section 7.6.

Finally, the idea of using an approximation algorithm for the Lagrangean relaxation of an optimization problem, and performing a parametric search to satisfy the constraint while preserving the approximation ratio, has a rich history in approximation algorithms. It is the method of choice for network design problems when there is a hard bound on the resource allocation constraint, for instance, $k$-medians [18] and $k$-MST [7]. We extend this technique to deal with the hard constraint on the rate of transmissions, and our results constitute the first application of this technique to policy design. This technique yields threshold-based reward policies, which suggests that this technique will have interesting connections to the retirement-based index policies [14] for multi-armed bandit problems – these connections will be explored in future work.

## 7.2 Proof of Lemma 7.3

We first prove that $\{\beta\}$ constitutes a feasible solution to for $\text{LPU}\textsc{nSat}(\epsilon)$. Since $\mathcal{S}_{\sigma_+} \le \lambda(1+\epsilon) < \mathcal{S}_{\sigma_-}$, $\beta_{\sigma_+} \in [0,1)$ and $\beta_{\sigma_-} \in (0,1]$. Finally, note that $\sum_{\sigma\in\Pi} \beta_\sigma = 1$ and $\sum_{\sigma\in\Pi} \beta_\sigma \mathcal{S}_\sigma = \lambda(1+\epsilon)$. The result follows.

We now prove that $\sum_{\sigma\in\{\sigma_{\mathcal{L}^+}, \sigma_{\mathcal{L}^-}\}} \beta_\sigma \mathcal{G}_\sigma \ge c(1-\epsilon) Q^*(\epsilon)$. Note that since $\sigma_+$ $c-$approximates $\text{LPL}\textsc{agrange}(\epsilon, \mathcal{L}^+)$ and $0 \le c \le 1$ and $\sum_{\sigma\in\Pi} \beta^*(\epsilon)_\sigma = 1$,

$$\mathcal{G}_{\sigma_+} + \mathcal{L}^+\left(\lambda(1+\epsilon) - \mathcal{S}_{\sigma_+}\right) \ge c\left[\sum_{\sigma\in\Pi} \beta^*(\epsilon)_\sigma \mathcal{G}_\sigma + \mathcal{L}^+\left(\lambda(1+\epsilon) - \sum_{\sigma\in\Pi} \beta^*(\epsilon)_\sigma \mathcal{S}_\sigma\right)\right] \tag{3}$$

And likewise for $\mathcal{L}^-$,

$$\mathcal{G}_{\sigma_-} + \mathcal{L}^-\left(\lambda(1+\epsilon) - \mathcal{S}_{\sigma_-}\right) \ge c\left[\sum_{\sigma\in\Pi} \beta^*(\epsilon)_\sigma \mathcal{G}_\sigma + \mathcal{L}^-\left(\lambda(1+\epsilon) - \sum_{\sigma\in\Pi} \beta^*(\epsilon)_\sigma \mathcal{S}_\sigma\right)\right] \tag{4}$$

Since $\{\beta^*(\epsilon)\}$ is a feasible of $\text{LPU}\textsc{nSat}(\epsilon)$ , $\sum_{\sigma\in\Pi} \beta^*(\epsilon)_\sigma \mathcal{S}_\sigma = \lambda(1+\epsilon)$. Thus the terms $\left(\lambda(1+\epsilon) - \sum_{\sigma\in\Pi} \beta^*(\epsilon)_\sigma \mathcal{S}_\sigma\right)$ can be removed from the respective right hand sides of the equations 3 and 4. We now multiply equation 3 by $\alpha$ and equation 4 by $1-\alpha$ and add the resulting equations. The right hand side of the sum evaluates to $c \sum_{\sigma\in\Pi} \beta^*(\epsilon)_\sigma \mathcal{G}_\sigma = cQ^*(\epsilon)$ . Since $\alpha\mathcal{S}_{\sigma_+} + (1-\alpha)\mathcal{S}_{\sigma_-} = \lambda(1+\epsilon)$, the left hand side becomes $\alpha\mathcal{G}_{\sigma_+} + (1-\alpha)\mathcal{G}_{\sigma_-} - (\mathcal{L}^+ - \mathcal{L}^-)(1-\alpha)\left(\lambda(1+\epsilon) - \mathcal{S}_{\sigma_-}\right)$ . Thus, we have

$$\alpha\mathcal{G}_{\sigma_+} + (1-\alpha)\mathcal{G}_{\sigma_-} - (\mathcal{L}^+ - \mathcal{L}^-)(1-\alpha)\left(\lambda(1+\epsilon) - \mathcal{S}_{\sigma_-}\right) \ge cQ^*(\epsilon)$$

Now since $0 \leq \lambda(1+\epsilon) \leq \mathcal{S}_{\sigma_-} \leq 1$, $-1 \leq (1-\alpha)(\lambda(1+\epsilon)-\mathcal{S}_{\sigma_-}) \leq 0$. Thus, since $0 \leq (\mathcal{L}^+ - \mathcal{L}^-) \leq c\epsilon Q^*(\epsilon)$ we have

$$\alpha \mathcal{G}_{\sigma_+} + (1-\alpha)\mathcal{G}_{\sigma_-} \geq cQ^*(\epsilon) - (\mathcal{L}^+ - \mathcal{L}^-) \geq c(1-\epsilon)Q^*(\epsilon)$$

The result follows since $\beta_{\sigma_+} = \alpha$ and $\beta_{\sigma_-} = 1-\alpha$.

### 7.3 Threshold policies for $c$-approximating LPLAGRANGE$(\epsilon, \mathcal{L})$

We first generalize the definition for $H_{u,\ell}$ as follows.

**Definition 7.6.** *Let $H_{u,\ell,x} = \phi$ for all $u \geq K$. For each $\ell$, starting from $u = K-1$, down to $u = w_\ell$, recursively, define $H_{u,\ell,x} = \left\{ i | i \notin \bigcup_{v:v>u} H_{v,\ell,x}, \text{ and } \tilde{r}_i[u] - \frac{c_i}{\tilde{p}_i[u]} > \max(\tilde{r}_\ell[0], r_{u-1}, x) \right\} \setminus \{\ell\}$. Assume that $c_i/\tilde{p}_i[u] = \infty$ when $\tilde{p}_i[u] = 0$. Let $w_{\ell,x} = \min_u\{u : u \geq 0, r_u > \max(\tilde{r}_\ell[0], x)\}$. If $r_u \leq \max(\tilde{r}_\ell[0], x)$ for all $u$, $w_{\ell,x} = K$.*

RESERVEBKUP$(\ell, x)$

**(Probing Process:)**
$u = K-1$
While $u \geq w_{\ell,x}$ and the highest state of a probed channel is lower than $u$,
  probe channels in $H_{u,\ell,x}$ in non-increasing order of $\tilde{r}_j[u] - c_j/\tilde{p}_j[u]$.
  $u \to u-1$

**(Selection Process:)**
Consider the channel $j$ in the highest state $y$ among all probed channels.
  (If no channel is probed, $j = -1$.)
  If $\max(r_y, \tilde{r}_\ell[0]) < x$, do not transmit. If $\max(r_y, \tilde{r}_\ell[0]) \geq x$, transmit in $j$ if $r_y \geq \tilde{r}_\ell[0]$ and transmit in $\ell$ otherwise.

Note that RESERVEBKUP$(\ell, x)$ is similar to RESERVEBKUP$(\ell)$ except that RESERVEBKUP$(\ell, x)$ selects a transmission threshold, $x$, apriori, and transmits only if a probed channel is in state $j$ or a higher state such that $r_j \geq x$ or if the probability of success of the backup channel $\ell$ is not lower than $x$. Thus, RESERVEBKUP$(\ell, x)$, probes only those channels for which the expected rewards conditioned on being in states $k$ and above, where $r_k > \max(\tilde{r}_\ell[0], x)$, exceed the probing cost.

**Definition 7.7.** *Let* BESTRESERVEBKUP$(x)$ *be the* RESERVEBKUP$(\ell, x)$ *for that $\ell$ for which it attains the maximum gain among all $\ell \in \{0, \ldots, n\}$, and $\sigma_x$ denote the decision tree of* BESTRESERVEBKUP$(x)$

In the next section, we prove that BESTRESERVEBKUP$(x)$ optimally solves $T(x)$ for $K = 2$ and 2/3-approximates $T(x)$ for $K > 2$. Lemma 7.4 follows since BESTRESERVEBKUP$(x)$ can be computed in $O(n^2K)$ time.

### 7.4 Proof of Lemma 7.4

The proof relies on the following Generalized Structure Theorem which is similar to the Structure Theorem 5.1.

**Theorem 7.7** (Generalized Structure Theorem)**.** *Consider the* SATURATED ALTERED REWARD $(x)$ *system described in Definition 7.5. There exists an optimum policy in this system that uses a unique backup channel whenever it transmits in a backup channel. The backup channel is used at the end of one $\leq i$ path.*

*Proof.* If no optimum policy transmits in a backup channel, the theorem clearly holds. Now, suppose there exists an optimum policy OPT1 that transmits in a backup channel at the end of some path in its decision tree. Observe that lemma 5.5 holds in this system. The proof is the same as that in the original system. If OPT1 does not probe any channel, then the theorem follows as well. Let OPT1 probe a channel. Clearly, then, OPT1 probes a channel $j$ at the root node, say $m$, of its decision tree. Let $i$ be the highest state of $j$ for which OPT1 transmits in a backup channel somewhere downstream of $m$. Then, by lemma 5.5, there exists another optimum policy OPT2 for which the decision tree rooted at $m$, and hence the overall decision tree, is (a) a $\leq i$ tree (b) uses a channel, say $\ell$, as a backup at the end of the $\leq i$ path in the tree and (c) uses $\ell$ as a backup whenever it uses a backup. The theorem follows. □

*Proof.* (of Lemma 7.4)

Consider the SATURATED ALTERED REWARD $(x)$ system described in Definition 7.5. The set of decision trees in this system is $\Pi$, irrespective of $x$. Note that the gain of any $\sigma \in \Pi$ in this system is $\mathcal{G}_\sigma - x\mathcal{S}_\sigma$ and depends on $x$. Let $\sigma_x^*$ be a policy that attains the maximum gain in this system (and hence solves problem $T(x)$), $F = \mathcal{G}_{\sigma_x} - x\mathcal{S}_{\sigma_x}$ and $BEST = \mathcal{G}_{\sigma_x^*} - x\mathcal{S}_{\sigma_x^*}$. We need to prove that $F = BEST$ for $K = 2$ and $F \geq (2/3)BEST$ for $K > 2$. Lemma 7.4 follows since BESTRESERVEBKUP$(x)$ can be computed in $O(n^2K)$ time.

In this system, for each $x$, multiple $\sigma$ may maximize the gain. The generalized structure theorem (Theorem 7.7) shows that for each $x$ at least one $\sigma_x^*$ uses a unique backup channel whenever a backup channel is used for transmission. We therefore consider a $\sigma_x^*$ that uses a unique backup, say $\ell$. Let $\sigma_x^*$ use $\ell$ as backup with probability $\alpha$. Let $R = \sum_{i:r_i \geq x} p_{i\ell}(r_i - x)$ and $T = \sum_{i:r_i < x} p_{i\ell}(x - r_i)$.

Let $\mathcal{P}'(j)$ be the set of all decision trees in $\Pi$ that use no channel other than $j$ as a backup and do not probe $j$. Note that policies in $\mathcal{P}'(0)$ never transmit in any backup channel. Using similar arguments as in the proof of Theorem 5.2 (which holds for both $K = 2$ and $K > 2$), it can be shown that in this system RESERVEBKUP$(j, x)$ attains the maximum gain among all policies in $\mathcal{P}'(j)$. Thus, $F$ is the maximum gain attained in this system by any policy in $\cup_{j=0}^n \mathcal{P}'(j)$.

Let $K = 2$. We now prove that $\sigma_x^* \in \cup_{j=0}^n \mathcal{P}'(j)$. It follows that $F \geq BEST$. Let $x \geq r_1$. Now, $\sigma_x^*$ does not transmit in any channel. Thus, $\sigma_x^* \in \mathcal{P}'(0)$. The result follows. Now, let $x < r_1$. Let $\sigma_x^* \notin \cup_{j=0}^n \mathcal{P}'(j)$. Given Theorem 7.7, the above happens only when $\sigma_x^*$ probes a channel $j$ in one path and uses it as a backup in another path. We now rule this out. Now, if a probed channel is in the highest state, state 1, $\sigma_x^*$ transmits in that channel. Thus, $\sigma_x^*$ consists of only one path, say $P$, and some other links that originate from $P$. Each of these links correspond to the case that a probed channel is in state 1 and leads to a leaf node at which $\sigma_x^*$ transmits in the probed channel. Thus, $\sigma_x^*$ can transmit in a backup channel only at the end of $P$, but then it can not have probed the channel in $P$, and hence does not probe the channel in any other path as well. The result follows.

Next, let $K > 2$. Now, construct a policy $\sigma_1$ that is similar to $\sigma_x^*$ except that whenever $\sigma_x^*$ uses $\ell$ as a backup, $\sigma_1$ does not transmit. Thus $\sigma_1$ attains a gain of at least BEST - $\alpha(R - T)$. Also, $\sigma_1 \in \mathcal{P}'(0)$. Thus, its gain is upper bounded by $F$. Thus $F \geq BEST - \alpha(R - T)$. Now, consider the policy that transmits in $\ell$ every slot without probing any channel. This policy is in $\mathcal{P}'(\ell)$ and attains a gain of $R - T$. Thus, $F \geq R - T$. It follows that

$$(1 + \alpha)F \geq BEST. \tag{5}$$

Next, construct another policy $\sigma_2$ that is similar to $\sigma_x^*$ except that $\sigma_2$ never probes $\ell$; instead wherever $\sigma_x^*$ probes $\ell$, $\sigma_2$ follows the same course of actions as $\sigma_x^*$ does after discovering $\ell$ in state 0. The gain of $\sigma_2$ is at least $BEST - (1 - \alpha)(R - c_\ell)$, because if $\ell$ was in state $i$ such that $r_i < x$, $\sigma_x^*$ will never transmit in $\ell$. Also, $\sigma_2 \in \mathcal{P}'(\ell)$. Therefore, $F \geq BEST - (1 - \alpha)(R - c_\ell)$. Now consider another policy $\sigma_3$ that probes $\ell$ and subsequently transmits only if $\ell$ is in a state $i$ such that $r_i \geq x$; $\sigma_3$ neither probes nor transmits in any other channel. Clearly, $\sigma_3$ attains a gain of $R - c_\ell$. Also, $\sigma_3 \in \mathcal{P}'(0)$. Thus, $F \geq R - c_\ell$. Combining the last two equations,

$$(2-\alpha)F \geq BEST. \tag{6}$$

Adding (5) and (6), we get $F \geq (2/3)BEST$. The result follows. □

## 7.5 Proof of Lemma 7.5

We now describe how the parameters $\mathcal{L}^+, \mathcal{L}^-$ and $\sigma_+, \sigma_-$ are selected.

**Definition 7.8.** *Let* THRESHOLD *be an array consisting of* $n+K+2$ *elements which are* $-1, 2, r_0, \ldots, r_{K-1}$ *and* $\tilde{r}_1[0], \ldots, \tilde{r}_n[0]$ *sorted in an increasing order.*

Note that the decision tree $\sigma_x$ of BESTRESERVEBKUP$(x)$ is the same for all $x \in$ (THRESHOLD$[i]$, THRESHOLD$[i+1]$). Thus, THRESHOLD is the collection of thresholds $x$ corresponding to different values of $\sigma_x$.

**Definition 7.9.** *For* $1 \leq i \leq n+K+1$ *let* $\hat{\sigma}_i$ *be* $\sigma_x$ *(i.e.,* BESTRESERVEBKUP $(x)$*) for* $x =$ THRESHOLD$[i]$.

**Lemma 7.8.** *For* $\epsilon \in (0, 1/\lambda - 1)$*, there exists an* $i$ *such that* $\mathcal{S}_{\hat{\sigma}_i} > \lambda(1+\epsilon) \geq \mathcal{S}_{\hat{\sigma}_{i+1}}$.

*Proof.* Now, $\mathcal{S}_{\sigma_x} = 1$ for $x \leq 0$ and $\mathcal{S}_{\sigma_x} = 0$ for $x > r_{K-1}$. Thus, since THRESHOLD$[1] = -1$ and THRESHOLD$[n+K+2] > r_{K-1}$, $\mathcal{S}_{\hat{\sigma}_1} = 1 > \lambda(1+\epsilon)$ and $\mathcal{S}_{\hat{\sigma}_{n+K+1}} = 0 \leq \lambda(1+\epsilon)$. The result follows since $\mathcal{S}_{\hat{\sigma}_i} \geq \mathcal{S}_{\hat{\sigma}_{i+1}}$ for each $i$. □

*Proof.* (of Lemma 7.5) Let $i^*$ be the $i$ found by Lemma 7.8, and $\Delta = \min(\frac{2\epsilon Q^*(\epsilon)}{3}$, (THRESHOLD$[i^*+1]-$ THRESHOLD$[i^*])/2)$.

If $\mathcal{S}_{\sigma_x} \leq \lambda(1+\epsilon)$, for $x \in$ (THRESHOLD$[i^*]$, THRESHOLD$[i^*+1]$), $\mathcal{L}^+ =$ THRESHOLD$[i^*] + \Delta$, and $\mathcal{L}^-$ =THRESHOLD$[i^*]$. If $\mathcal{S}_{\sigma_x} > \lambda(1+\epsilon)$, for $x \in$ (THRESHOLD$[i^*]$, THRESHOLD$[i^*+1]$), $\mathcal{L}^+$ =THRESHOLD$[i^*+1]$, and $\mathcal{L}^- =$ THRESHOLD$[i^*+1] - \Delta$. Now, $\sigma^+ = \sigma_{\mathcal{L}^+}, \sigma^- = \sigma_{\mathcal{L}^-}$.

Since $\mathcal{S}_{\hat{\sigma}_{i^*}} > \lambda(1+\epsilon) \geq \mathcal{S}_{\hat{\sigma}_{i^*+1}}$, in both cases $\mathcal{L}^+, \mathcal{L}^-$ and $\sigma_+, \sigma_-$ satisfy the properties of lemma 7.3.

Note that we need to compute $\sigma_x$ for $O(n+K)$ values of $x$. Thus, the guarantee on the computation time follows since $\sigma_x$ can be computed in $O(n^2K)$ time (lemma 7.4). □

## 7.6 Algorithm Summary

We now summarize the design of the stable policy, UNSATAPPROX$(\epsilon)$ that attains a gain of $\frac{1-\epsilon}{1+\epsilon}G_U$ for $K=2$ and $(2/3)\frac{1-\epsilon}{1+\epsilon}G_U$ for $K>2$ (Theorem 7.6). Recall that $\sigma_x$ is BESTRESERVEBKUP$(x)$, $\hat{\sigma}_i$ is $\sigma_x$ for $x =$ THRESHOLD$[i]$ (Definition 7.8).

UNSATAPPROX$(\epsilon)$

1. Compute $\mathcal{L}^+, \mathcal{L}^-$ as in the proof of lemma 7.5, and let $\sigma^+, \sigma^-$ denote BESTRESERVEBKUP$(\mathcal{L}^+)$ and BESTRESERVEBKUP$(\mathcal{L}^-)$ respectively.
2. In each busy slot, use $\sigma^+$ with probability $\alpha$ and $\sigma^-$ with probability $1-\alpha$, where $\alpha = \frac{\mathcal{S}_{\sigma_-} - \lambda(1+\epsilon)}{\mathcal{S}_{\sigma_-} - \mathcal{S}_{\sigma_+}}$.

# 8 Conclusions

We have presented a simple model for studying the information acquisition and exploitation trade-off at a single wireless node, when the multiple available channels are multi-state, and the channel distributions and information acquisition costs could be different. We presented a general solution framework based on exploiting the structure of the optimal policy, and by using Lagrangean relaxations to simplify the space of approximately optimal solutions. We believe these techniques will have wider applicability, in particular when we consider the multiple node scenario.